\documentstyle[preprint,aps,prc,psfig]{revtex}
\tighten
\begin{document}

\title{Improved lower bounds for the ground-state energy of many-body systems}
\author{D. Van Neck, Y. Dewulf, and M. Waroquier}
\address{Laboratory of Theoretical Physics, Ghent University, 
Proeftuinstraat 86, B-9000 Gent, Belgium}
\maketitle
\begin{abstract}
New lower bounds for the binding energy of a quantum-mechanical system 
of interacting particles are presented. The new bounds are expressed 
in terms of two-particle quantities and improve the conventional bounds 
of the Hall-Post type. They are constructed by considering not only 
the energy in the two-particle system, but also the structure of the pair 
wave function. We apply the formal results to various numerical examples, 
and show that in some cases dramatic improvement over the existing bounds 
is reached.  
\end{abstract}

\section{Introduction}\label{section1}
One of the central problems in quantum many-body physics is to find 
the energy of a system of $A$ particles interacting with given 
two-body potentials. Variational techniques yield an upper bound to the 
exact energy. The determination of a strict lower bound can then provide a 
natural and useful complement. 

In recent years there has been renewed interest in deriving such lower bounds, 
mostly in connection with quark models in hadron spectroscopy \cite{Bas1-2}, 
or the limits for Borromean phenomena in loosely bound systems 
\cite{Ric,Goy,Mos,Gri}. Earlier uses of lower bounds were focussed on 
thermodynamical considerations \cite{old1}, or the problem of stability in  
self-gravitating systems \cite{old2,old3}.

Up to now all lower bounds are based on the Hall-Post decomposition 
\cite{hp1,hp2} of the hamiltonian into two-body clusters, and subsequent 
application of the variational principle in two-body space. Several variants 
and extensions have been proposed, e.g.\ an optimal decomposition in case of 
three \cite{Bas3} or four \cite{Ben} unequal masses. The case of identical 
fermions was recently studied in \cite{Jui}.   
 
In its most useful form, a lower bound of the Hall-Post type is expressed in 
terms of the ground-state energy in a two-body system. Finding this 
two-body energy usually goes together with determining the wave function of 
the ground-state pair. 
We show in this paper that by using the structure of the pair wave 
function one can always improve the Hall-Post bound, and that the 
improvement is sometimes spectacular. 
Our results apply to boson and 
fermion systems, both with and without the presence of an external potential, 
and irrespective of the local or nonlocal character of the two-body 
potentials.  
We do restrict ourselves to systems of identical particles, leaving a study 
of the unequal mass case for future work. 

Loosely speaking, the Hall-Post bound implies that the ground state for 
$A$ identical particles is a superposition of $\frac{1}{2}A(A-1)$ pairs in the 
lowest-energy state of a modified hamiltonian. However, because of the 
correlated structure of such a pair, it is in general not possible 
to reach a pair occupation of $\frac{1}{2}A(A-1)$, except in the case of 
noninteracting 
bosons. For any given pair, there exists a maximal value for the  
occupancy that can be reached in an $A$-body state. The remainder of the 
$\frac{1}{2}A(A-1)$ pairs must then necessarily be in a state with higher 
energy. This allows the construction of a new lower bound, whose value is 
increased with respect to the Hall-Post case. 

The present paper is organized as follows. In 
Section~\ref{section2} we first establish notations, introduce cluster 
decompositions of the many-body hamiltonian, and rederive 
the Hall-Post inequalities. Next we point out in Section~\ref{section3} 
how the Hall-Post bound can be improved in the case of fixed-center bosonic 
systems. The equivalent case for fermions is treated in 
Section~\ref{section4}. The modifications that enter when considering 
self-bound (translationally invariant) systems are discussed in 
Section~\ref{section5}. Finally, we apply the formal results to 
various numerical examples, which are collected in Section~\ref{section6}. 
Section~\ref{section7} contains a global summary and points out some 
remaining problems.

\section{General remarks on cluster decompositions of the hamiltonian for 
a system of identical particles}\label{section2}
\subsection{Notational conventions}\label{section2.1}
We restrict ourselves to systems of identical particles and follow the 
notational conventions of \cite{prc}. In particular, single-particle 
coordinates are generically written as $x_1 ,x_2 ,\dots $, and should be 
regarded as spatial coordinates in $d$ dimensions. Spin or isospin degrees 
of freedom are not explicitly mentioned, but may be assumed to be included 
in the $x_i$ dependence of the wave functions and operators. 
  
For a system of $A$ particles we consider a hamiltonian $\hat{H}_A [\alpha]$ 
which is a sum of one-body and two-body operators with variable relative 
weight,
\begin{equation}
\label{eq:1}
\hat{H}_A [\alpha](x_1 ,\dots ,x_A )=\alpha
\sum_{i_1 =1,\dots ,A} t(x_{i_1})+
\sum_{i_1 < i_2 =1,\dots ,A}v(x_{i_1} ,x_{i_2}).
\end{equation}
Using a combinatorial identity we can rewrite $\hat{H}_A [\alpha =1]$, for 
any $N$ with $2\leq N\leq A$, as
\begin{equation}
\label{eq:2}
\hat{H}_A [1](x_1 ,\dots ,x_A )=\frac{1}{\tiny    
\left(\begin{array}{c}A-2\\N-2\end{array}\right) }
\sum_{i_1 < \dots < i_N =1,\dots ,A}\hat{H}_N [\frac{N-1}{A-1}]
(x_{i_1 },\dots ,x_{i_N }) .
\end{equation}

The hamiltonian $\hat{H_N}[\frac{N-1}{A-1}]$ for clusters of $N$ particles 
is related to the original hamiltonian $\hat{H_A}[1]$ for $A$ particles, 
but has different relative weights for its one-body and two-body components 
(or, equivalently: a different coupling strength of the two-body 
interaction).  
We can now make the weights equal again by absorbing the difference in an 
additional one-body term. As shown in Section~\ref{section2.2}, this leads to 
sumrules for 
the energy which are generalizations of the well-known Koltun sumrule 
\cite{Kol}. Alternatively, we can keep the 
different coupling strength in the $N$-particle hamiltonian. 
This will lead to another set of sumrules, derived in 
Section~\ref{section2.3}, from 
which the Hall-Post inequalities immediately follow. 

One of the ingredients in these sumrules are the spectroscopic factors 
related to the removal of particles from the $A$-particle system in its 
ground state. These are defined as follows\footnote{
The definitions in this Section may look tedious or superfluous, but 
will allow in later Sections a unified treatment for fixed-center and 
self-bound systems.}.

The $A$-particle ground state 
$\Psi_{0(A)}[1]$ of $\hat{H}_A [1]$ can always be expanded in terms of the 
complete orthonormal set of $N$-particle eigenstates
$\Psi_{\nu (N)}[\alpha]$ of $\hat{H}_N [\alpha]$,
\begin{equation}
\label{eq:3}
\Psi_{0(A)}[1](x_1 ,\dots ,x_A )=
{\tiny \left(\begin{array}{c}A\\N\end{array}\right) }^{-1/2}
\sum_{\nu (N)}\Psi_{\nu (N)}[\alpha](x_1 ,\dots ,x_N )
\psi_{\nu (N)}[\alpha](x_{N+1} ,\dots ,x_A ).
\end{equation}
The expansion coefficients $\psi_{\nu (N)}[\alpha]$ are the overlap 
functions between  the $\Psi_{\nu (N)}[\alpha]$ and $\Psi_{0(A)}[1]$,
\begin{eqnarray}
\label{eq:4}
\psi_{\nu (N)}[\alpha](x_{N+1},\dots ,x_{A})=&&\\
&&{\tiny \left(\begin{array}{c}A\\N\end{array}\right) }^{1/2}
\int dx_1 \dots dx_N \Psi^*_{\nu (N)}[\alpha](x_1 ,\dots ,x_N )
\Psi_{0(A)}[1](x_1 ,\dots ,x_A ) ,\nonumber
\end{eqnarray}
and their normalization yields the corresponding spectroscopic factor
$S_{\nu (N)}[\alpha]$,
\begin{equation}
\label{eq:5}
S_{\nu (N)}[\alpha]=\int dx_{N+1}\dots dx_A |\psi_{\nu (N)}[\alpha]
(x_{N+1},\dots ,x_{A})|^2 ,
\end{equation}
that is, the squared amplitude for removing $A-N$ particles from 
$\Psi_{0(A)}[1]$ and ending up in $\Psi_{\nu (N)}[\alpha]$. 

Obviously, the spectroscopic factors are positive, and the completeness of the 
set $\Psi_{\nu (N)}[\alpha]$ implies the sumrule 
\begin{equation}
\label{eq:6}
\sum_{\nu (N)} S_{\nu (N)}[\alpha]= 
{\tiny \left(\begin{array}{c}A\\N\end{array}\right) } .
\end{equation}

Note that the 
$S_{\nu (N)}[\alpha]$ can also be interpreted as the 
{\em occupation number} of the state $\Psi_{\nu (N)}[\alpha]$ in the 
ground state $\Psi_{0(A)}[1]$, since Eq.(\ref{eq:5}) can be rewritten as 
\begin{eqnarray}
\label{eq:new} 
S_{\nu (N)}[\alpha]&=&\int dx_{1}\dots dx_N dx'_{1}\dots dx'_N 
\Psi_{\nu (N)}^*[\alpha](x_{1},\dots ,x_N )
\rho^{(N)}(x_{1},\dots ,x_N ;x'_{1},\dots ,x'_N )\nonumber\\
&&\Psi_{\nu (N)}[\alpha](x'_{1},\dots ,x'_N ), 
\end{eqnarray}
in terms of the $N$-body density matrix
\begin{eqnarray}
&&\rho^{(N)}(x_{A-N+1},\dots ,x_A ;x'_{A-N+1},\dots ,x'_A )=
{\tiny \left(\begin{array}{c}A\\N\end{array}\right) }
\int dx_{1}\dots dx_{A-N}\\&&
\Psi_{0(A)}^* [1](x_1 ,\dots ,x_{A-N} ,x'_{A-N+1},\dots ,x'_A )
\Psi_{0(A)}[1](x_1 ,\dots ,x_{A-N} ,x_{A-N+1},\dots ,x_A ) .\nonumber
\end{eqnarray}

The above equations hold for fixed-center systems, in which the one-body part  
of the hamiltonian contains an external potential. For self-bound 
systems the one-body part is purely kinetic, and the two-body part 
depends on relative coordinates only, i.e.\  
$v(x_1 ,x_2 )\equiv v(x_1 -x_2 )$. Since the intrinsic eigenstates of the 
hamiltonian are translationally invariant, some modifications \cite{prc} 
are needed.

The intrinsic kinetic energy is obtained by subtracting the 
center-of-mass from the total kinetic energy, 
\begin{equation}
\hat{T'}_A =-\frac{1}{2m}\sum_{i_1 =1,\dots ,A}\partial^2_{i_1} 
+\frac{1}{2Am}(\sum_{i_1 =1,\dots ,A}\partial_{i_1} )^2 
=-\frac{1}{2Am}\sum_{i_1 \leq i_2 =1,\dots ,A}
(\partial_{i_1}-\partial_{i_2})^2 ,
\end{equation} 
and is seen to behave as a two-body operator with an $A$-dependent 
coupling strength. 
Defining the intrinsic hamiltonian with variable coupling strength as
\begin{equation}
\hat{H'}_A [\alpha ]= \alpha \hat{T'}_A +
+\sum_{i_1 < i_2 =1,\dots ,A}v(x_{i_1} -x_{i_2}),
\end{equation}
the decomposition (\ref{eq:2}) gets modified to
\begin{equation}
\label{eq:7}
\hat{H'}_A [1](x_1 ,\dots ,x_A )=\frac{1}{\tiny   
\left(\begin{array}{c}A-2\\N-2\end{array}\right) }
\sum_{i_1 < \dots < i_N =1,\dots ,A}\hat{H'}_N [\frac{N}{A}]
(x_{i_1 },\dots ,x_{i_N }) .
\end{equation}

The $A$-particle ground-state wave function can be expanded in a 
complete orthonormal set of intrinsic $N$-particle eigenstates \cite{prc}, 
the self-bound equivalent of Eq.(\ref{eq:3}) being 
\begin{eqnarray}
\label{eq:11}
\Psi_{0(A)}[1](x_1 ,\dots ,x_A )&=&
{\tiny \left(\begin{array}{c}A\\N\end{array}\right) }^{-1/2}
\sum_{\nu (N)}\Psi_{\nu (N)}[\alpha](x_1 ,\dots ,x_N )\nonumber\\
&&\times\psi_{\nu (N)}[\alpha](x_{N+1}-R_N ,\dots ,x_A -R_N ),
\end{eqnarray}
where $R_N =\frac{1}{N}\sum_{i=1}^N x_i$; the overlap functions
are defined as
\begin{eqnarray}
\label{eq:12}
\psi_{\nu (N)}[\alpha](x_{N+1},\dots ,x_{A})&=&
{\tiny \left(\begin{array}{c}A\\N\end{array}\right) }^{1/2}
\int dx_1 \dots dx_N \delta (R_N )\Psi^*_{\nu (N)}[\alpha](x_1 ,\dots ,x_N )
\nonumber\\&&\times 
\Psi_{0(A)}[1](x_1 ,\dots ,x_A ) ,
\end{eqnarray}
whereas Eqs.(\ref{eq:5},\ref{eq:6}) remain unchanged.

The spectroscopic factor $S_{\nu (N)}$, as defined by Eq.(\ref{eq:5}), can 
again be viewed as the 
occupation number of the intrinsic $N$-particle state $\Psi_{\nu (N)}$ 
in the $A$-particle ground state, but this occupation number now also includes 
a summation over all possible states of $N$-particle center-of-mass 
motion, i.e.\ using 
\begin{equation}
\int dK \frac{\exp{(-iKR_N )}}{(2\pi )^{d/2}}\frac{\exp{(iKR'_N )}}
{(2\pi )^{d/2}} =\delta (R_N -R'_N ) , 
\end{equation}
we can express the spectroscopic factor as 
\begin{eqnarray}
\label{snew} 
S_{\nu (N)}[\alpha]&=&\int dx_{1}\dots dx_N dx'_{1}\dots dx'_N 
\delta (R_N -R'_N ) 
\Psi_{\nu (N)}^*[\alpha](x_{1},\dots ,x_N )\\&&
\rho^{(N)}(x_{1},\dots ,x_N ;x'_{1},\dots ,x'_N )
\Psi_{\nu (N)}[\alpha](x'_{1},\dots ,x'_N ).\nonumber 
\end{eqnarray}
The $N$-body density matrix for a self-bound system, defined as
\begin{eqnarray}
&&\rho^{(N)}(x_{A-N+1},\dots ,x_A ;x'_{A-N+1},\dots ,x'_A )=
{\tiny \left(\begin{array}{c}A\\N\end{array}\right) }
\int dx_{1}\dots dx_{A-N}\delta (R_{A-N}) \\&&
\Psi_{0(A)}^* [1](x_1 ,\dots ,x_{A-N} ,x'_{A-N+1},\dots ,x'_A )
\Psi_{0(A)}[1](x_1 ,\dots ,x_{A-N} ,x_{A-N+1},\dots ,x_A ) ,\nonumber
\end{eqnarray}
is the proper extension of the one-body density matrix for translationally 
invariant systems in \cite{prc,prcp}.

\subsection{Generalized Koltun sumrules}\label{section2.2}

For fixed-center systems the decomposition (\ref{eq:2}) can be 
reshuffled so as to have the same hamiltonian 
in $A$-particle and $N$-particle space, at the expense of an 
additional one-body term,
\begin{equation}
\label{eq:12bis}
\hat{H}_A [1](x_1 ,\dots ,x_A )=
\frac{1}{\tiny
\left(\begin{array}{c}A-2\\N-2\end{array}\right)
}
\sum_{i_1 < \dots < i_N =1,\dots ,A}
\hat{H}_N [1](x_{i_1 },\dots ,x_{i_N })
-\frac{A-N}{N-1}\sum_{i_1 =1,\dots ,A}t(x_{i_1}) .
\end{equation}

The expectation value of (\ref{eq:12bis}) in 
the $A$-particle ground state, expanded according to Eq.(\ref{eq:3}),  
can be worked out using Eqs.(\ref{eq:4},\ref{eq:5}) and reads
\begin{equation}
\label{eq:13}
E_{0(A)}[1]=\frac{N}{A+N-1}\left\{
T_{0(A)}-\frac{1}{\tiny \left(\begin{array}{c}A-2\\N-1\end{array}\right)}
\sum_{\nu (N)}(E_{\nu (N)}[1]-E_{0(A)}[1])S_{\nu (N)}[1] \right\} .
\end{equation}
Here the ground-state energy has been expressed in terms of the expectation 
value $T_{0(A)}$ of the one-body field and the 
first energy-weighted moment of the 
distribution of $(A-N)$-particle removal strength (or mean removal energy).
This is a generalization of the familiar Koltun sumrule \cite{Kol} for $N=A-1$,
\begin{equation}
E_{0(A)}[1]=\frac{1}{2}\left\{
T_{0(A)}-\sum_{\nu (A-1)}\left(E_{\nu (A-1)}[1]-E_{0(A)}[1]\right)
S_{\nu (A-1)}[1]\right\},
\end{equation}
where the distribution of single-particle removal strength is 
experimentally accessible through single-particle (SP) knock-out 
reactions \cite{Kel,cmo16}.

For self-bound systems we can [in analogy to Eq.(\ref{eq:12bis})] restore 
in Eq.(\ref{eq:7}) the original hamiltonian at the expense of an 
additional (intrinsic) kinetic term,
\begin{equation}
\label{eq:10}
\hat{H'}_A [1](x_1 ,\dots ,x_A )=
\frac{1}{\tiny
\left(\begin{array}{c}A-2\\N-2\end{array}\right)
}
\sum_{i_1 < \dots < i_N =1,\dots ,A}
\hat{H'}_N [1](x_{i_1 },\dots ,x_{i_N })
-\frac{A-N}{N}\hat{T'}_A . 
\end{equation}
The ground-state expectation value of Eq.(\ref{eq:10}), evaluated by means of 
Eqs.(\ref{eq:11},\ref{eq:12}), then reads 
\begin{equation}
E_{0(A)}[1]=\frac{N}{A+N-1}\left\{
\frac{N-1}{N}T'_{0(A)}-\frac{1}
{\tiny \left(\begin{array}{c}A-2\\N-1\end{array}\right)}
\sum_{\nu (N)}(E_{\nu (N)}[1]-E_{0(A)}[1])S_{\nu (N)}[1] \right\}.
\end{equation}
The result for the case of $N=A-1$, 
\begin{equation}
E_{0(A)}[1]=\frac{1}{2}\left\{\frac{A-2}{A-1}
T'_{0(A)}-\sum_{\nu (A-1)}(E_{\nu (A-1)}[1]-E_{0(A)}[1])
S_{\nu (A-1)}[1]\right\},
\end{equation}
is the Koltun sumrule for self-bound systems (including the correct 
recoil factor for the kinetic energy \cite{prc,Die}).
 
\subsection{Hall-Post inequalities}\label{section2.3}

The expectation value of Eq.(\ref{eq:2}), with the modified coupling strength 
in the $N$-particle hamiltonian, can be evaluated in the same way 
as Eq.(\ref{eq:13}),
\begin{equation} 
\label{eq:18}
E_{0(A)}[1] =\frac{1}
{\tiny \left(\begin{array}{c}A-2\\N-2\end{array}\right)}
\sum_{\nu (N)} E_{\nu (N)}[\frac{N-1}{A-1}]S_{\nu (N)}[\frac{N-1}{A-1}].
\end{equation}
This expression forms the basis for deriving the lower bounds considered 
in the next Section. At the simplest level, combining Eq.(\ref{eq:18}) 
with the trivial inequality resulting from the spectroscopic sumrule 
(\ref{eq:6}),  
\begin{equation}
\label{trivial}
0\leq S_{\nu (N)}[\frac{N-1}{A-1}] \leq 
{\tiny \left(\begin{array}{c}A\\N\end{array}\right)} ,
\end{equation}
immediately leads to inequalities 
of the Hall-Post type
\begin{equation}
\label{eq:31}
E_{0(A)}[1] \geq\frac{A(A-1)}{N(N-1)}
E_{0(N)}[\frac{N-1}{A-1}] .
\end{equation}

For self-bound systems the $A$-particle ground-state energy can be likewise 
evaluated,  from the expectation value of Eq.(\ref{eq:7}), in terms of 
intrinsic $N$-particle energies and occupation numbers, and reads
\begin{equation} 
\label{eqself}
E_{0(A)}[1] =\frac{1}
{\tiny \left(\begin{array}{c}A-2\\N-2\end{array}\right)}
\sum_{\nu (N)} E_{\nu (N)}[\frac{N}{A}]S_{\nu (N)}[\frac{N}{A}].
\end{equation}
The Hall-post inequalities
\begin{equation}
\label{hpinsb}
E_{0(A)}[1] \geq\frac{A(A-1)}{N(N-1)}
E_{0(N)}[\frac{N}{A}] ,
\end{equation}
then follow immediately from Eq.(\ref{trivial}).
 
\section{Improved lower bounds for fixed-center bosonic 
systems}\label{section3}

\subsection{Derivation of new lower bounds}\label{section3.1}

From now on we concentrate on the most relevant case $N=2$, and try to 
derive lower bounds for the $A$-particle ground-state energy in terms of 
two-body quantities.
At present, such a lower bound is given by the Hall-Post inequality 
(\ref{eq:31}),
\begin{equation}
\label{eq:32}
E_{0(A)}[1] \geq
{\tiny \left(\begin{array}{c}A\\2\end{array}\right)}
E_{0(2)}[\frac{1}{A-1}] ,
\end{equation}
which is 
usually derived by applying a variational principle 
in two-body space \cite{Ric}.
As was shown in Section~\ref{section2.3}, the Hall-Post inequality is actually 
an {\em equality} when expressed in terms of two-body energies and occupations,
\begin{equation}
\label{eq:33}
E_{0(A)}[1] =
\sum_{\nu (2)} E_{\nu (2)}[\frac{1}{A-1}]S_{\nu (2)}[\frac{1}{A-1}] .
\end{equation}

In order to simplify notations we drop the coupling strength dependence 
in this Section, as it will be fixed to $\alpha=1$ for $A$-particle and 
$\alpha = \frac{1}{A-1}$ for two-particle quantities. 
It is clear from Eqs.(\ref{eq:32},\ref{eq:33}) that in the traditional 
lower bound the distribution of pair strength $S_{\nu (2)}$ is approximated by 
concentrating all strength in the two-body ground state, i.e. by the 
distribution 
\begin{equation}
S_{\nu (2)}\approx\delta_{0\nu}
{\tiny \left(\begin{array}{c}A\\2\end{array}\right)} ,
\end{equation}
which exhausts the sumrule (\ref{eq:6}). 

This assumption can only be realistic for weakly correlated systems. For 
a non-interacting Bose system [$v\equiv 0$ in Eq.(\ref{eq:1})]  
it holds exactly. The uncorrelated eigenstates are product-type 
wave functions,  
\begin{equation}
\Psi^{\mbox{\scriptsize unc}}_{0(A)}(x_1 ,\dots ,x_A )
=\prod_{i=1,\dots ,A}\chi (x_i ),\,\,\,\,\,
\Psi^{\mbox{\scriptsize unc}}_{0(2)}(x_1 ,x_2 )=\chi (x_1 )\chi (x_2 ),
\end{equation}
where $\chi (x)$ is the SP eigenstate 
of the one-body hamiltonian $t(x)$ corresponding to the lowest 
energy $\epsilon_0$. 
As a consequence, $S_{0(2)}^{\mbox{\scriptsize unc}}=A(A-1)/2$, and the 
Hall-Post lower bound coincides with the exact result, 
\begin{equation}
E_{0(A)}^{\mbox{\scriptsize unc}}=A\epsilon_0 = 
{\tiny \left(\begin{array}{c}A\\2\end{array}\right)}
\frac{1}{A-1}2\epsilon_0 =
{\tiny \left(\begin{array}{c}A\\2\end{array}\right)}
E_{0(2)}^{\mbox{\scriptsize unc}} .
\end{equation}
For strongly correlated systems we expect that the occupation 
$S_{0(2)}$ can 
deviate substantially from $A(A-1)/2$, and the Hall-Post lower bound will be 
far from the exact result.  

In order to improve on this situation we need to take into account 
correlations in the structure of the many-boson 
eigenstates. 
As it is more convenient to work in second quantization, we 
define $\varphi_{\nu (2)}^{\dagger}$ as the creation operator for the two-body 
eigenstate $\Psi_{\nu (2)}$, i.e.\ 
\begin{equation}
\mid\Psi_{\nu (2)}\rangle = \varphi^{\dagger}_{\nu (2)}\mid \rangle .
\end{equation}
The occupation $S_{\nu (2)}$, as defined 
by Eq.(\ref{eq:5}) or Eq.(\ref{eq:new}), is then 
simply written as
\begin{equation} 
S_{\nu (2)}=\langle\Psi_{0(A)}\mid\varphi^{\dagger}_{\nu (2)}
\varphi^{\mbox{}}_{\nu (2)}\mid\Psi_{0(A)}\rangle .
\end{equation}

We now consider the {\em upper bound} 
$S_{\nu (2)}^{\mbox{\scriptsize max}}$ 
for $S_{\nu (2)}$, 
\begin{equation}
\label{upper}
S_{\nu (2)}^{\mbox{\scriptsize max}}=\max_{\Psi_A } \left\{\langle\Psi_A \mid
\varphi^{\dagger}_{\nu (2)}\varphi^{\mbox{}}_{\nu (2)}\mid\Psi_A
\rangle\right\} ,
\end{equation}
where the maximum is taken with respect to all normalized $A$-boson 
eigenstates.  
This upper bound only requires knowledge of the two-body state 
$\Psi_{\nu (2)}$;  its explicit construction is pointed out in 
Section~\ref{section3.2}.

Since $S_{\nu (2)}^{\mbox{\scriptsize max}}$ is better than the trivial 
upper bound (\ref{trivial}), the following inequality holds,
\begin{equation}
S_{\nu (2)}\leq S_{\nu (2)}^{\mbox{\scriptsize max}} \leq 
{\tiny \left(\begin{array}{c}A\\2\end{array}\right)}.
\end{equation}
In combination with Eq.(\ref{eq:33}) this results in a sequence 
of new lower bounds ${\cal L}_{\mu}$, with $\mu =0,1,\dots$\,, 
for the ground-state energy, 
\begin{equation}
\label{lower}
E_{0(A)}\geq {\cal L}_{\mu} = \sum_{\nu =0}^{\mu -1}
S_{\nu (2)}^{\mbox{\scriptsize max}}
E_{\nu (2)}+\left\{
{\tiny \left(\begin{array}{c}A\\2\end{array}\right)}
-\sum_{\nu =0}^{\mu -1}S_{\nu (2)}^{\mbox{\scriptsize max}}\right\}
E_{\mu (2)} .
\end{equation}
Here the two-body states are assumed to be ordered according to 
increasing energy, $E_{0(2)}\leq E_{1(2)}\leq\dots$.

The optimal lower bound in this sequence is given by the largest value  
${\cal L}_{\tilde{\mu}}=\max_{\mu}({\cal L}_{\mu})$, 
where $\tilde{\mu}$ is determined by 
\begin{equation}
\sum_{\nu =0}^{\tilde{\mu} -1}S_{\nu (2)}^{\mbox{\scriptsize max}}< 
{\tiny \left(\begin{array}{c}A\\2\end{array}\right)}\leq
\sum_{\nu =0}^{\tilde{\mu}}S_{\nu (2)}^{\mbox{\scriptsize max}} .
\end{equation}

The inequalities in Eq.(\ref{lower}) constitute  the principal result 
in this paper, and several remarks are in order:\newline 
(a) The conventional Hall-Post bound of Eq.(\ref{eq:32}) coincides with 
${\cal L}_0$.\newline
(b) ${\cal L}_1$ is {\em always} a better bound 
than ${\cal L}_0$, since 
\begin{equation}
{\cal L}_1 -{\cal L}_0 =\left\{  
{\tiny \left(\begin{array}{c}A\\2\end{array}\right)}-
S_{0 (2)}^{\mbox{\scriptsize max}}\right\}[E_{1(2)}-E_{0(2)}]\geq 0.
\end{equation}
In most of the cases we studied ${\cal L}_1$ is the optimal bound 
${\cal L}_{\tilde{\mu}}$.\newline
(c) If only a finite number $n_b$ of discrete levels is present in the 
two-body spectrum, then Eq.(\ref{lower}) holds for $0\leq\mu\leq n_b$, 
where $E_{n_b}$ can be taken equal to zero.\newline
(d) Eq.(\ref{lower}) holds without any symmetries of the underlying 
hamiltonian. If these are present, they can of course be used to refine 
the lower bound, e.g.\ an energy level $E_{\nu (2)}$ appearing in 
Eq.(\ref{lower}) can correspond 
to a $d_{\nu}$-fold degenerate multiplet with eigenstates 
$\varphi_{\nu (2),\mu},\,\mu=1,\dots ,d_{\nu}$. If the quantum numbers of the 
ground state $\Psi_{0(A)}$ are known, it will usually be possible to 
determine the maximal {\em joint} occupation number of the multiplet,
\begin{equation}
P_{\nu (2)}^{\mbox{\scriptsize max}}=
\max_{\Psi_A } \left\{\langle\Psi_A \mid\sum_{\mu}
\varphi^{\dagger}_{\nu (2),\mu}\varphi^{\mbox{}}_{\nu (2),\mu}\mid\Psi_A
\rangle\right\} ,
\end{equation}
where the variation is made over all $A$-particle states with the same 
symmetry properties as $\Psi_{0(A)}$. Using 
$P_{\nu (2)}^{\mbox{\scriptsize max}}$ in the evaluation of Eq.(\ref{lower}) 
will result in a better bound, because 
$P_{\nu (2)}^{\mbox{\scriptsize max}}\leq \sum_{\mu}
S_{\nu (2),\mu}^{\mbox{\scriptsize max}}$. An example of this will be given 
in Section~\ref{section6}.\newline
(e) Finally, doing better than ${\cal L}_{\tilde{\mu}}$ would 
require e.g.\ an optimalization of the simultaneous occupation 
for the two-body ground and first excited state, involving a determination of 
\begin{equation}
\max_{\Psi_A} \langle\Psi_A \mid
E_{0(2)}\varphi^{\dagger}_{0(2)}\varphi^{\mbox{}}_{0(2)}+
E_{1(2)}\varphi^{\dagger}_{1(2)}\varphi^{\mbox{}}_{1(2)}
\mid\Psi_A\rangle .
\end{equation}
This appears to be a far more complicated problem than the determination of 
$S_{\nu (2)}^{\mbox{\scriptsize max}}$.

\subsection{Construction of maximal pair occupancies}\label{section3.2}

Introducing the set of its natural orbitals $\chi_{\alpha}(x)$, a general  
two-boson state $\Psi_{(2)}$ can be written as (see Appendix~\ref{app1.1}),
\begin{equation}
\Psi_{(2)} (x_1 ,x_2) = \sum_{\alpha} x_{\alpha} \chi_{\alpha}(x_1 )
\chi_{\alpha}(x_2 ), 
\end{equation}
where the $x_{\alpha}$ are real and positive, and 
$\sum_{\alpha}x_{\alpha}^2 =1$.
In second quantized form this reads  
\begin{equation}
\mid\Psi_{(2)}\rangle=\frac{1}{\sqrt{2}}\sum_{\alpha} x_{\alpha} 
\left( c^{\dagger}_{\alpha}\right)^2 \mid\rangle =
\varphi_{(2)}^{\dagger} \mid\rangle,
\end{equation}
where $c^{\dagger}_{\alpha}$ is the creation operator for the one-boson state 
$\chi_{\alpha}(x)$. 

The maximal pair occupation $S_{(2)}^{\mbox{\scriptsize max}}$ of 
$\Psi_{(2)}$, defined according to Eq.(\ref{upper}), is equal to the 
largest eigenvalue of the following hermitian eigenvalue problem 
in $A$-boson space, 
\begin{equation}
\label{eig}
\varphi_{(2)}^{\dagger}\varphi_{(2)}^{\mbox{}}\mid\Psi_A \rangle=
\lambda\mid\Psi_A \rangle .
\end{equation}

The corresponding eigenvalue problem for fermions was recently solved by 
Pan et al.\cite{Pan}, in the context of 
a generalized pairing problem. The method in \cite{Pan}, which involves an 
infinite-dimensional algebra, can be easily adapted to 
bosonic systems. Here we only state the final result; for completeness 
the derivation is given in Appendix~\ref{app2}.

Let $s=0$ for $A$ even, $s=1$ for $A$ odd, and $a=(A-s)/2$.
The largest eigenvalue of $\varphi_{(2)}^{\dagger}\varphi_{(2)}^{\mbox{}}$ 
is among the eigenvalues $\lambda$ in the maximally paired subspace 
(see Appendix~\ref{app2}). These  are given by
\begin{equation}
\label{eigenvalue}
\lambda =1+2sx_{\beta}^2 +4\sum_{k=1,\dots ,a-1}\frac{1}{y_k}, 
\end{equation}
in terms of the solutions of a set of $a-1$ nonlinear equations 
in $a-1$ variables $(y_1 ,\dots ,y_{a-1})$,  
\begin{equation}
\label{difficult}
\frac{1}{4}\sum_{\alpha}\frac{x_{\alpha}^2 (1+2s\delta_{\alpha\beta})}
{1-y_k x_{\alpha}^2}=\frac{1}{y_k} +\sum_{l=1,\dots ,a-1;l\neq k}
\frac{1}{y_k -y_l} ,\,\,\,\, k=1,\dots ,a-1 .
\end{equation}
The index $\beta$ which appears for odd $A$ is arbitrary 
(see Appendix~\ref{app2}).

The system of equations (\ref{difficult}) allows to determine the 
maximal occupancy $S_{(2)}^{\mbox{\scriptsize max}}$ for general $A$. 
For small values of $A$ it can be rewritten in a much simpler form:
\begin{eqnarray}
(A=3)&\,\rightarrow\,& \lambda=1+2x_{\beta}^2 ,\nonumber\\
(A=4)&\,\rightarrow\,& \sum_{\alpha}\frac{x_{\alpha}^2}
{\lambda -1-4x^2_{\alpha}} = 1 ,\nonumber\\
(A=5)&\,\rightarrow\,& \sum_{\alpha}\frac{x_{\alpha}^2 
(1+2\delta_{\alpha\beta})}
{\lambda -1-4x^2_{\alpha}-2x^2_{\beta}} = 1 .\nonumber
\end{eqnarray}

The maximal occupation of a pair state depends on the structure of 
$\Psi_{(2)}$, i.e.\ on the distribution of the $x_{\alpha}$. The two 
extreme cases are: (a) uncorrelated, and (b) ``maximally correlated''. 
In the uncorrelated limit (a) only one of the $x_{\alpha}$ is non-zero, 
resulting in $S_{(2)}^{\mbox{\scriptsize max}}=A(A-1)/2$. 
The latter limit (b) has a flat distribution, i.e.\ assuming 
that there are 
$\Omega$ SP states, then $x_{\alpha}=\frac{1}{\sqrt{\Omega}}$. 
This case, corresponding to a schematic boson pairing force in a single 
degenerate shell, can be treated analytically since the algebra reduces to 
SU(2). The resulting maximal eigenvalue 
is $S_{(2)}^{\mbox{\scriptsize max}}= a(1+\frac{2(a-1+s)}{\Omega})$. 

In summary, the maximal 
eigenvalue in $A$-boson space obeys
\begin{equation} 
a\leq S_{(2)}^{\mbox{\scriptsize max}}\leq 
{\tiny \left(\begin{array}{c}A\\2\end{array}\right)} ,
\end{equation}
where the upper limit corresponds to an uncorrelated wave function, and the 
lower limit is reached as the large-$\Omega$ limit of a maximally correlated 
wave function.

\section{Modifications for fixed-center fermionic systems}\label{section4}

The basic inequalities (\ref{lower}) derived in the previous Section 
still hold for fermions. However, the boundaries 
for the allowed pair occupation numbers are completely different.

The natural orbitals for a  two-fermion state come in associated pairs 
$(\chi_{\alpha}(x),\chi_{\bar{\alpha}}(x))$, and a  
general two-fermion state $\Psi_{(2)}$ can be written as 
(see Appendix~\ref{app1.2}),
\begin{equation}
\Psi_{(2)} (x_1 ,x_2) = \sum_{\alpha >0} x_{\alpha} 
[\chi_{\alpha}(x_1 )\chi_{\bar{\alpha}}(x_2 )-\chi_{\bar{\alpha}}(x_1 )
\chi_{\alpha}(x_2 )], 
\end{equation}
where the sum runs over distinct pairs.
The $x_{\alpha}$ are real, positive for $\alpha > 0$, and with 
$x_{\bar{\alpha}}=-x_{\alpha}$, normalized as $\sum_{\alpha}x_{\alpha}^2 =1$.
In second quantized form this reads  
\begin{equation}
\mid\Psi_{(2)}\rangle=\sqrt{2}\sum_{\alpha >0} x_{\alpha} 
c^{\dagger}_{\alpha}c^{\dagger}_{\bar{\alpha}}\mid\rangle =
\varphi_{(2)}^{\dagger} \mid\rangle .
\end{equation}

The maximal pair occupation $S_{(2)}^{\mbox{\scriptsize max}}$ of 
$\Psi_{(2)}$ is again equal to the 
largest eigenvalue $\lambda$ of an eigenvalue problem 
in $A$-fermion space, 
\begin{equation}
\label{eigfer}
\varphi_{(2)}^{\dagger}\varphi_{(2)}^{\mbox{}}\mid\Psi_A \rangle=
\lambda\mid\Psi_A \rangle .
\end{equation}

For fermion systems we can use directly the results in \cite{Pan}.
Let $s=0$ for $A$ even and $s=1$ for $A$ odd; and $a=(A-s)/2$.
We have to solve the set of $a-1$ non-linear equations in $a-1$ variables
$(y_1 ,\dots ,y_{a-1})$, 
\begin{equation}
-\frac{1}{2}\sum_{\alpha > 0}\frac{x_{\alpha}^2 (1-s\delta_{\alpha\beta})}
{1-y_k x_{\alpha}^2}=\frac{1}{y_k} +\sum_{l=1,\dots ,a-1;l\neq k}
\frac{1}{y_k -y_l} ,\; k=1,\dots ,a-1.
\end{equation}
Then the relevant eigenvalues of (\ref{eigfer}) 
corresponding to the subspace of 
maximally paired states are given by 
\begin{equation}
\lambda =1-2sx_{\beta}^2 -4\sum_{k=1,\dots ,a-1}\frac{1}{y_k}
\end{equation}

The simplified secular equations for small $A$ read,
\begin{eqnarray}
(A=3)&\,\rightarrow\,& \lambda=1-2x_{\beta}^2 ,\nonumber\\
(A=4)&\,\rightarrow\,& 2\sum_{\alpha > 0}\frac{x_{\alpha}^2}
{\lambda -1+4x^2_{\alpha}} = 1 ,\nonumber\\
(A=5)&\,\rightarrow\,& 2\sum_{\alpha > 0}\frac{x_{\alpha}^2 
(1-\delta_{\alpha\beta})}
{\lambda -1+4x^2_{\alpha}+2x^2_{\beta}} = 1 .\nonumber
\end{eqnarray}

It is again instructive to consider the two limiting cases in the structure 
of $\Psi_{(2)}$. In the uncorrelated case only one of the coefficients 
$x_{\alpha}$ is non-zero. This corresponds to a two-body Slater determinant, 
and $S_{(2)}^{\mbox{\scriptsize max}}=1$.  
In the maximally correlated case, where all coefficients are equal, the 
coefficients become $\mid x_{\alpha}\mid=\frac{1}{\sqrt{\Omega}}$, if 
there are $\Omega$ SP states. This is equivalent to the well-known problem 
of a schematic fermion pairing force in a single degenerate shell, with 
$S_{(2)}^{\mbox{\scriptsize max}}=a(1-\frac{2(a-1+s)}{\Omega})$. 

In the large-$\Omega$ limit we then find that 
the maximal pair occupation in $A$-fermion space obeys  
\begin{equation} 
1\leq S_{(2)}^{\mbox{\scriptsize max}}\leq a .
\end{equation}
Note the different role of correlations for boson and fermion systems: for 
boson systems correlations decrease the maximal pair occupation compared 
to the uncorrelated case, for fermion systems they increase it.

From these bounds on $S_{(2)}^{\mbox{\scriptsize max}}$ it follows that 
the Hall-Post lower bound (\ref{eq:32}) will never be satisfactory 
for fermion systems. Even without knowing the structure of the two-body 
eigenstates, it can be replaced by the better bound
\begin{equation} 
E_{0(A)}\geq a\sum_{\nu (2)}E_{\nu (2)} ,
\end{equation}
where $\nu =0,\dots ,2(a-1+s)$.
Unfortunately, this bound does not in general become the exact result 
in the limit of noninteracting fermions. For a noninteracting fermionic 
system the new bound 
(\ref{lower}) would yield ${\cal L}_{\tilde{\mu}}=\sum_{\nu (2)}E_{\nu (2)}$, 
where $\nu=0,\dots, \frac{1}{2}A(A-1)-1$; hence the new bound, though better, 
would still not be exact, the reason being that the $\frac{1}{2}A(A-1)$ 
two-particle energies $E_{ij}=\epsilon_i +\epsilon_j$ made with the 
$A$ lowest SP energies $\epsilon_i$, are not necessarily the 
lowest two-particle energies. 
  
\section{Modifications for self-bound systems}\label{section5}
In analogy to the treatment in the previous Sections we try to derive 
lower bounds for the intrinsic $A$-body ground-state energy in terms of 
the two-body wave functions and energies of relative motion, by considering  
Eq.(\ref{eqself}) with $N=2$, 
\begin{equation}
E_{0(A)}[1]=\sum_{\nu (2)}E_{\nu (2)}[\frac{2}{A}]S_{\nu (2)}[\frac{2}{A}].
\end{equation}
We again drop the dependence on the coupling strength, since it will remain 
fixed at $\alpha =1$ for the $A$-body, and at $\alpha =\frac{2}{A}$ for the 
two-body quantities.

Apart from the different couplings there are no differences with the 
fixed-center case, and the basic set of inequalities Eqs.(\ref{lower}) is 
still valid.
The novel complication lies in deriving an upper bound 
$S_{\nu (2)}^{\mbox{\scriptsize max}}$ for the pair 
occupation of a relative pair wave function $\Psi_{(2)}$ in an intrinsic 
$A$-particle wave function $\Psi_{(A)}$.  
Mathematically this boils down to 
finding the absolute maximum of the pair occupation in Eq.(\ref{eq:5}) or 
Eq.(\ref{snew}),
\begin{eqnarray}
\label{self}
S_{(2)}
&=&{\tiny \left(\begin{array}{c}A\\N\end{array}\right) }  
\int dx_1\, dx_2\, dx'_1\, dx'_2\, \delta (R_2 )\delta (R'_2 )
\Psi_{(2)}^* (x_1 ,x_2 )\Psi_{(2)}(x'_1 ,x'_2 )
\nonumber\\&& 
\int dx_3 \dots dx_A \Psi_{(A)} (x_1 ,x_2, x_3 ,\dots, x_A)
\Psi^*_{(A)} (x'_1 ,x'_2 ,x_3 , \dots, x_A)
\end{eqnarray}
by varying $\Psi_{(A)}$ in the space of translationally invariant 
wave functions of the correct (anti)symmetry. 

The problem of spectroscopic factors and occupation numbers in self-bound
systems is a difficult one (see \cite{prc}), and 
we did not succeed in finding a general solution to this problem.
The case $A=3$ is tractable, however, since it can be transformed into an 
eigenvalue equation in SP coordinate space.
We neglect (iso)spin degrees of freedom and only consider here cases where 
the spatial part of the wave function is totally symmetric ($\eta=1$) or 
antisymmetric ($\eta=-1$), though the results can probably be extended to 
cases of mixed spatial symmetry \cite{Jui}. 

Introducing Jacobi coordinates $a=x_1 -x_2$ and $b=x_3 -R_2 $, a 
general three-body wave function $\Psi_{(3)}$ can be written as 
\begin{equation}
\Psi_{(3)}(x_1 ,x_2 ,x_3 )\equiv \tilde{\Psi}_{(3)} (a,b)
=f(a,b)+\eta f(b+\frac{a}{2},\frac{3a}{4}-\frac{b}{2})
+f(b-\frac{a}{2},-\frac{3a}{4}-\frac{b}{2}), 
\end{equation}
in terms of a function $f(a,b)=\eta f(-a,b)$.
The pair wave function is simply 
\begin{equation}
\Psi_{(2)}(x_1 ,x_2 )\equiv g(a)=
\eta g(-a) .
\end{equation}

In terms of these quantities, the pair occupation (\ref{self}) for $A=3$ 
is rewritten as
\begin{eqnarray}
\label{eq:59}
S_{(2)}&=&
3\int db\, da\, da'\, f^* (a',b)g^* (a)\left[
g(a')\tilde{\Psi}_{(3)}(a,b)+\eta 
g(\frac{a'}{2} +b)\tilde{\Psi}_{(3)}(a,\frac{3a'}{4}-
\frac{b}{2})\right. \nonumber\\
&&\left. +g(-\frac{a'}{2} +b)\tilde{\Psi}_{(3)}(a,-\frac{3a'}{4}
-\frac{b}{2})\right] , 
\end{eqnarray}
and the maximum must be taken with respect to all $f(a,b)$ having a fixed 
normalization 
\begin{equation}
\int dx_1\, dx_2\, dx_3\, \delta (R_3)\mid\Psi_{(3)}(x_1 ,x_2 , x_3)\mid^2
=3\int da\, db\, f^* (a,b)\tilde{\Psi}_{(3)}(a,b) .
\end{equation}
Performing the variation leads to the secular equation
\begin{eqnarray}
\lambda \tilde{\Psi}_{(3)}(a',b)&=&
\int da\, g^* (a)\left[ 
g(a')\tilde{\Psi}_{(3)}(a,b)+\eta 
g(\frac{a'}{2} +b)\tilde{\Psi}_{(3)}(a,\frac{3a'}{4}
-\frac{b}{2})\right. \nonumber\\
&&\left. +g(-\frac{a'}{2} +b)\tilde{\Psi}_{(3)}(a,-\frac{3a'}{4}
-\frac{b}{2})\right].
\end{eqnarray}
Introducing the overlap function 
$G(b)=\int da\, g^* (a)\tilde{\Psi}_{(3)}(a,b)$ [see Eq.(\ref{eq:12})], 
the secular equation is transformed to a SP eigenvalue equation of a 
hermitian non-local operator,
\begin{equation}
\label{neweq}
(\lambda -1 )G(x)=2\eta\int dx'\left[
g^* (\frac{x}{2}+x')g(\frac{x'}{2} +x)\right]G(x') , 
\end{equation}
which can easily be solved numerically.

Although we cannot yet determine the maximal pair occupation for general $A$, 
we can still find a bound for $E_{0(A)}$ in terms of 
two-particle quantities that is better than the traditional Hall-Post bound  
Eq.(\ref{hpinsb}) for $N=2$. This is done simply by replacing in the Hall-Post 
lower bound (\ref{hpinsb}) for $N=3$,
\begin{equation}
\label{patch}
E_{0(A)}[\alpha=1]\geq \frac{A(A-1)}{6}E_{0(3)}[\frac{3}{A}],
\end{equation} 
the three-body ground-state energy $E_{0(3)}[\frac{3}{A}]$ by an improved 
lower bound obtained by solving Eq.(\ref{neweq}). 

\section{Numerical examples}\label{section6}
\subsection{Trapped boson system with pairing forces}\label{section6.1}
As a first example we consider a system of spinless bosons trapped in a 
(three-dimensional) harmonic-oscillator well, and interacting with a 
general monopole pairing force (see, e.g., Dukelsky et al. \cite{Duk}). 
The hamiltonian reads
\begin{equation}
\label{eq:57}
\hat{H}=\sum_N \epsilon_N \sum_{lm}c^{\dagger}_{Nlm}
c^{\mbox{}}_{Nlm}+gP^{\dagger}P,
\end{equation}
where $N=0,1,\dots$ is the harmonic oscillator (HO) quantum number 
and $l$ the orbital angular momentum. For convenience we remove the zero-point 
energy from the single-particle spectrum and put $\hbar\omega=1$, i.e. 
we take $\epsilon_N =N$.

The pair operator in Eq.(\ref{eq:57}) is 
\begin{equation}
P^{\dagger}= \sum_N w_N\sum_l \left(c^{\dagger}_{Nl}\cdot 
c^{\dagger}_{Nl} \right)=\sum_N w_N\sum_l (-1)^l \sqrt{2l+1}
\left[c^{\dagger}_{Nl}\otimes 
c^{\dagger}_{Nl} \right]^0_0 .
\end{equation} 
For the purely schematic pairing force, with constant $w_N\equiv w $, 
the system is exactly solvable  for any finite number of 
HO levels, as was demonstrated by Richardson\cite{Richardson}. 
However, the schematic force has some 
unrealistic features due to its implicit dependence on the 
degeneracy $D_N =(N+1)(N+2)/2$ of the HO shells\cite{Duk}. The interaction   
between boson pairs in $N$ and $N'$ levels is proportional 
to $\sqrt{D_N D_{N'}}$. For attractive pairing, e.g., this leads to    
occupations of the higher levels far exceeding those of the lower ones. 
This can be cured by taking $w_N = 1/\sqrt{D_N}$, which is the pairing force 
we will consider in our numerical examples. 

In order to have the same notation as in Section~\ref{section3.2}, we can go 
over to 
the natural basis for $P^{\dagger}$ by defining new SP states, 
\begin{eqnarray}
b^{\dagger}_{Nlm}&=&\frac{1}{\sqrt{2}}\mbox{i}^m 
\mbox{i}^{\frac{1}{2}(1-\mbox{sgn}(m))}
\left(c^{\dagger}_{Nlm}+\mbox{sgn}(m)c^{\dagger}_{Nl-m}\right)
,\;\mbox{if}\,m\neq 0 ,\nonumber\\
b^{\dagger}_{Nl0}&=&c^{\dagger}_{Nl0},
\end{eqnarray}
in terms of which $P^{\dagger}=\sum_{Nlm}w_N(b^{\dagger}_{Nlm})^2$. 

The construction in Eq.(\ref{lower}) of a lower bound ${\cal L}_{\mu}$ for 
the $A$-boson system requires first 
to solve the two-boson problem with the same hamiltonian (\ref{eq:57}) but 
modified SP energies $\epsilon'_{N} =\frac{N-1}{A-1}\epsilon_N$, and we 
briefly discuss its solution. The two-body eigenstates of the collective 
pairing type (involving all harmonic oscillator shells) can be written as 
\begin{equation}
\mid\Psi_{(2)}\rangle =\frac{1}{\sqrt{2}}\sum_{Nlm}x_{Nlm}
(b^{\dagger}_{Nlm})^2 \mid\rangle , 
\end{equation}
and have eigenenergies $E_{(2)}$ which are solutions of 
\begin{equation}
\label{eigcol}
\frac{1}{2g}=\sum_N \frac{w_N^2 D_N}{E_{(2)}-2\epsilon'_N}. 
\end{equation} 
The corresponding wave function is 
$x_{Nlm}\sim w_N /(E_{(2)}-2\epsilon'_N)$; these coefficients must be used in 
Eq.(\ref{difficult}) to determine the maximal occupation number of this 
two-boson state. The noncollective eigenstates have energies 
$E_{(2)}=\epsilon'_{N_1}+\epsilon'_{N_2}$, and keep their unperturbed 
(harmonic oscillator) structure, apart from the fact that for a level 
$E_{(2)}=2\epsilon'_{N}$ the pair wave function must be orthogonal to the 
zero-coupled pair $\sum_{lm}(b^{\dagger}_{Nlm})^2$. 
For $g<0$, the two-body ground-state is always collective, whereas the first 
excited state can be either collective or noncollective, depending on the 
interaction strength.  

\begin{figure}
\centering{\
\psfig{file=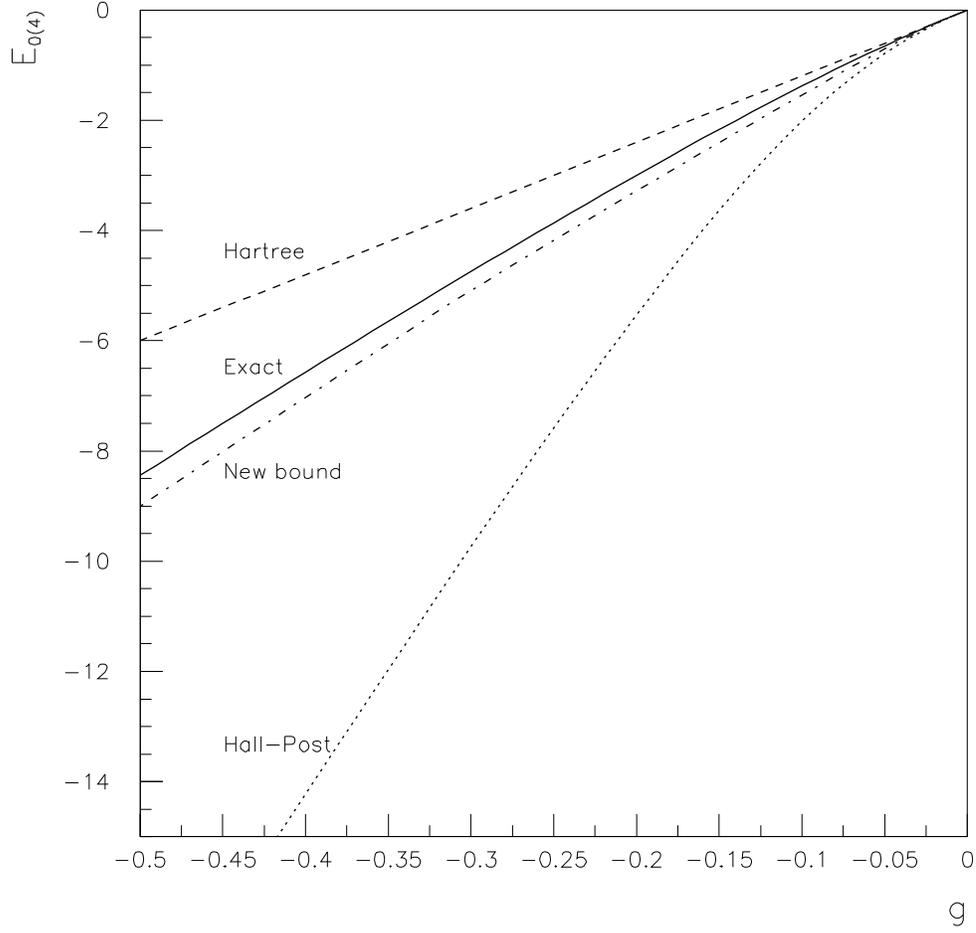,height=15cm}
\caption[fig1]{
Ground-state energy of four bosons with hamiltonian 
(\protect\ref{eq:57}), in four harmonic oscillator levels. Solid line: 
exact result $E_{0(4)}$ . Dashed line: Hartree upper bound $E^{H}_{(4)}$. 
Dotted line: Hall-Post 
lower bound ${\cal L}_0$. Dot-dashed line: improved lower bound 
${\cal L}_1$. 
}
\label{fig1}
}
\end{figure}

\begin{figure}
\centering{\
\psfig{file=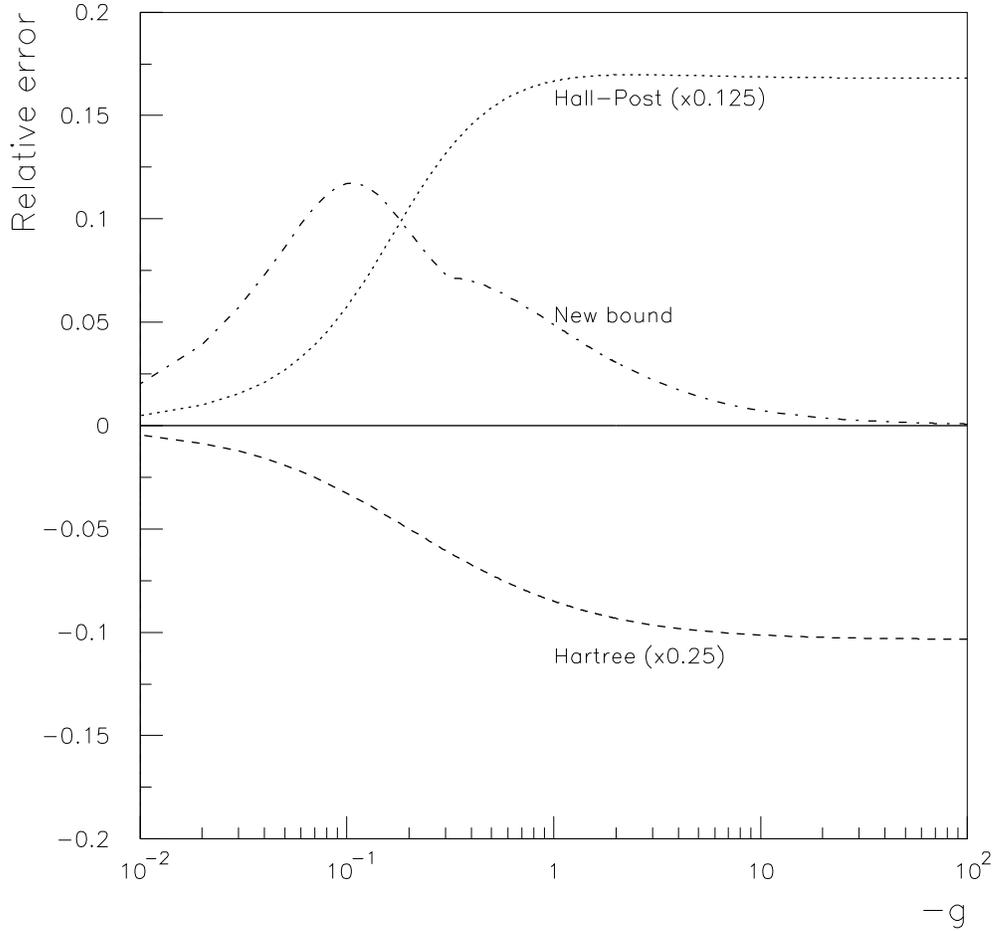,height=15cm}
\caption[fig2]{ Relative errors for the 
ground-state energy of four bosons with hamiltonian 
(\protect\ref{eq:57}), in four harmonic oscillator levels. Dashed line:
Hartree result  $(E^H_{(4)}-E_{0(4)})/E_{0(4)}$ [where $E_{0(4)}$ is the 
exact energy]. Dotted line: Hall-Post result 
$({\cal L}_0 -E_{0(4)})/E_{0(4)}$. Dot-dashed line: improved result 
$({\cal L}_1 -E_{0(4)})/E_{0(4)}$. 
}
\label{fig2}
}
\end{figure}

First we study a simple case of four bosons in four HO 
levels ($N=0,\dots ,3$), as the dimensionality is 
still sufficiently small to allow comparison with exact diagonalization. 
In Fig.~\ref{fig1} and Fig.~\ref{fig2} results are shown for 
attractive pairing.
The weak-coupling regime is displayed in Fig.~\ref{fig1}, where the 
exact energy is compared with the Hall-Post lower bound 
${\cal L}_0 =6E_{0(2)}$ and the new lower bound 
${\cal L}_1 =S^{\mbox{\scriptsize max}}_{0(2)}E_{0(2)}
+(6-S^{\mbox{\scriptsize max}}_{0(2)})E_{1(2)}$. 
We also plot, as an example of a simple upper bound, the energy $E^H_{(4)}$ of 
the Hartree solution, which is known exactly for this system. If the sequence 
of structure coefficients $w_N$ is decreasing with $N$, then 
the Hartree energy $E^H_{(A)}$ for general $A$ can be shown to equal 
$E^H_{(A)} /A =\epsilon_0 +(A-1)g$, if $g<g_c$, and 
$E^H_{(A)} /A =\epsilon_0 +(A-1)g_c (2-\frac{g_c}{g})$, if $g<g_c$, where the 
critical strength is $g_c=\frac{w_0 (\epsilon_1 - \epsilon_0)}{2(A-1)
(w_0 +w_1)}$.  

Both lower bounds coincide with the exact result as $g\rightarrow -0$. For 
more negative values of $g$, the Hall-Post bound quickly diverges from the 
exact result, whereas the improved lower bound follows the exact result quite 
closely. In fact, it is easy to see that 
the improved lower bound, in contrast to the Hall-Post one, becomes exact 
also in the strong-coupling limit, for all attractive pairing forces.
This is because, as $g\rightarrow -\infty$, the interaction term increasingly 
dominates over the external potential in the ground-state energy, and the 
lower bound ${\cal L}_1$ becomes exact for a separable hamiltonian 
$P^{\dagger}P$. Fig.~\ref{fig2}, where we show the relative error with respect 
to the exact energy, demonstrates this explicitely. The kink in 
${\cal L}_1$ at $g=-5/16\approx 0.3$ occurs because at this value of the 
coupling 
strength the first excited state of the $A=2$ system changes from 
a solution of Eq.(\ref{eigcol}) to the unpaired solution 
$\epsilon'_{0}+\epsilon'_{1}$. 

We checked that in all cases ${\cal L}_1$ is the optimal bound, i.e.\ 
$S^{\mbox{\scriptsize max}}_{0(2)}+S^{\mbox{\scriptsize max}}_{1(2)}>6$, 
by calculating $S^{\mbox{\scriptsize max}}_{1(2)}$ through  
Eq.(\ref{difficult}) for the second lowest solution of 
of Eq.(\ref{eigcol}), or, if the first excited state is the triplet 
$c^{\dagger}_{000}c^{\dagger}_{11\mu}\mid\rangle$, by realizing that 
the maximal joint occupation number of this triplet is equal to four in 
four-boson space.  

\begin{figure}
\centering{\
\psfig{file=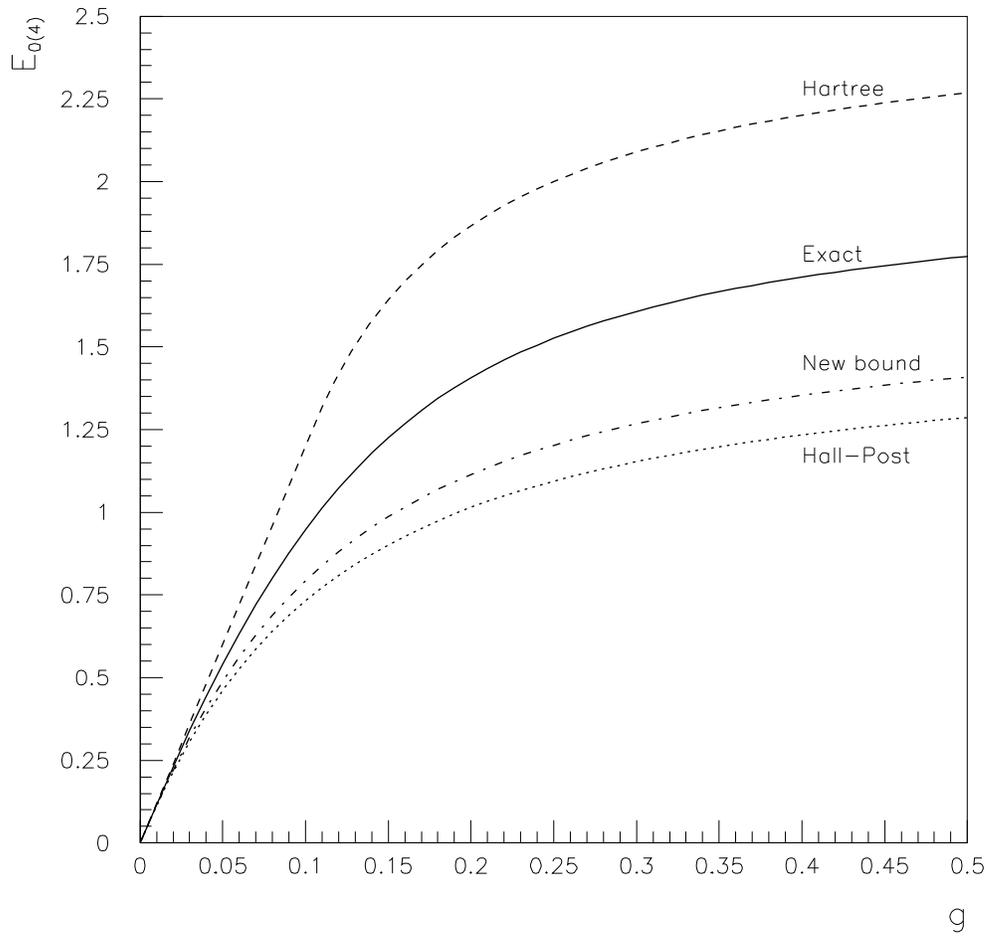,height=15cm}
\caption[fig3]{See caption Fig.~\protect\ref{fig1}. Case of repulsive 
pairing.
}
\label{fig3}
}
\end{figure}

The results for repulsive pairing ($g>0$), shown in Fig.~\ref{fig3}, are less 
impressive. In this case we lose the feature that as $g\rightarrow +\infty$ 
the two-body force dominates the ground-state energy; the system will 
simply tend to make pairs orthogonal to $P^{\dagger}$, and the one-body 
part of the hamiltonian can never be neglected. Of course, the new 
bound ${\cal L}_1$ (which is the optimal ${\cal L}_{\tilde{\mu}}$) is still 
better than the conventional ${\cal L}_0$. 

\begin{figure}
\centering{\
\psfig{file=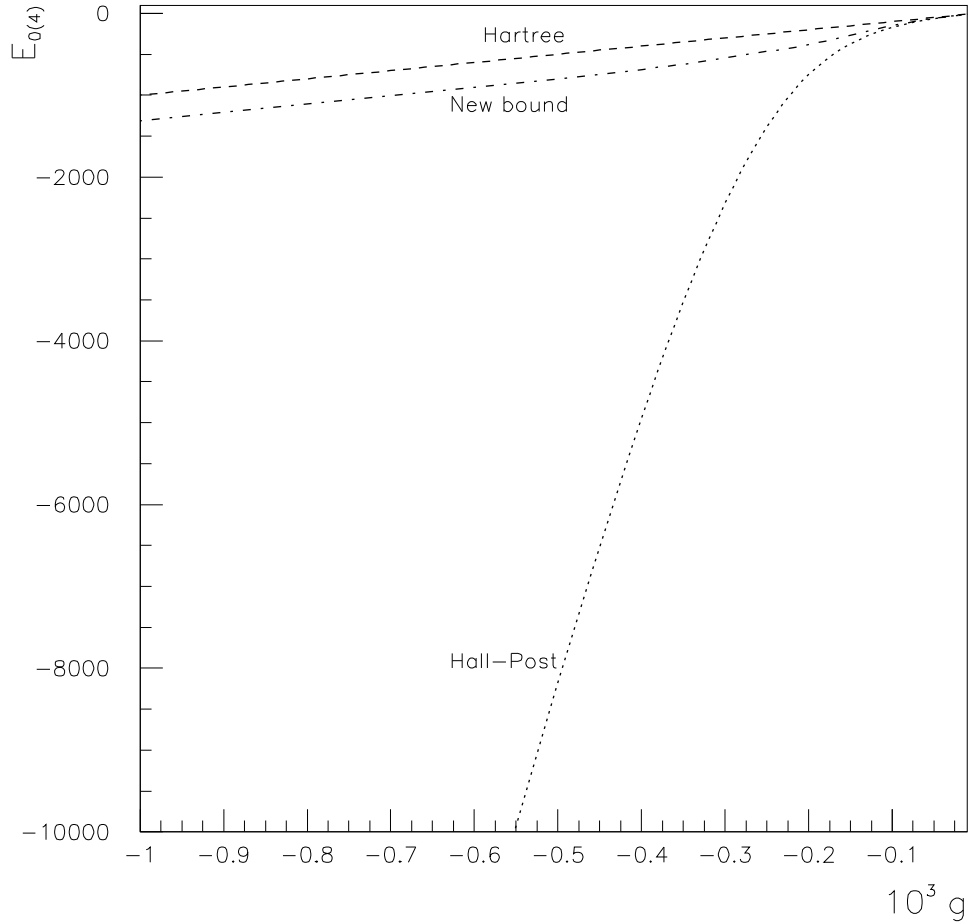,height=15cm}
\caption[fig4]{
Ground-state energy $E_{0(A)}$ for $A=1000$ bosons with hamiltonian 
(\protect\ref{eq:57}) in 50 harmonic oscillator levels. 
Dashed line: Hartree upper bound. Dotted line: Hall-Post 
lower bound ${\cal L}_0$. Dot-dashed line: improved lower bound 
${\cal L}_{\tilde{\mu}}$, where $\tilde{\mu}=1$ or $\tilde{\mu}=2$ (see text).
}
\label{fig4}
}
\end{figure}

Systems with a larger number of particles and/or shells can be similarly 
treated. Results for $1000$ bosons in $50$ harmonic oscillator main shells 
are shown in Fig.~\ref{fig4}, and the appreciable improvement of the new 
lower bound 
over the conventional one is again clear. For $g<g'$, 
where $g'\approx -2.6\times 10^{-4}$, we have ${\cal L}_2 >{\cal L}_1$ and 
the optimal lower bound ${\cal L}_{\tilde{\mu}}$ is given by ${\cal L}_2$ 
instead of ${\cal L}_1$; the difference between ${\cal L}_2$ and 
${\cal L}_1$ is marginal, however, certainly when compared with ${\cal L}_0$. 

\subsection{Bosons interacting with power-law potentials}\label{section6.2}

As an example of a self-bound system we consider $A$ spinless bosons in three 
dimensions, interacting with power-law potentials,  
\begin{equation}
\label{powlawham}
\hat{H}_A =-\frac{1}{2m}\sum_i {\bf \nabla}^2_i +g\,\mbox{sgn}(\beta )
\sum_{i_1 < i_2} \mid {\bf r}_i -{\bf r}_j \mid^{\beta} .
\end{equation}
Note that physically relevant potentials must have $\beta > -2$ to ensure 
an eigenvalue spectrum bounded from below.  

In accordance with Eq.(\ref{patch}), we try to derive lower bounds for 
the three-particle ground-state energy 
$E_{0(3)}[\frac{3}{A}]$, in terms of the solutions of the following relative 
pair hamiltonian,
\begin{equation}
\hat{H}'_2 = \left(\frac{3}{A}\frac{2}{3}\right)\left(
-\frac{1}{m}{\bf \nabla}^2\right) +g\,\mbox{sgn}(\beta )r^{\beta}.
\end{equation}
Scaling laws can be used to write the eigenenergies $E_{(2)}$ of 
$\hat{H}'_2$ as 
\begin{equation}
E_{(2)} = g\left(\frac{2}{Amg}\right)^{\frac{\beta}{\beta +2}}
\eta_{(2)}, 
\end{equation}
in terms of the eigenenergies $\eta_{(2)}$ of $-{\bf \nabla}^2 +
\mbox{sgn}(\beta )r^{\beta}$.

For the latter hamiltonian we determined numerically the $L^{\pi}=0^+$ 
ground-state energy $\eta_{0(2)}$ and wave function 
$\Psi_{0(2)}=g(r)\mbox{Y}_{00}$, as well as 
the energy $\eta_{1(2)}$ of the first excited (symmetric) state, which has 
$L^{\pi}=0^+$ for $\beta \leq 2$ and $L^{\pi}=2^+$ for $\beta \geq 2$. 

We also determined the maximal occupation 
$S_{0(2)}^{\mbox{\scriptsize max}}$ of the 
ground-state pair $\Psi_{0(2)}$ in $0^+$ three-boson space, which  
according to Eq.(\ref{neweq}) equals the largest 
eigenvalue $\lambda$ of
\begin{equation}
(\lambda -1)G(r)=2\int dr' {r'}^2 W(r,r')G(r').
\end{equation}  
The operator $W$ reads as 
\begin{equation}
\label{opera}
W(r,r')=\frac{1}{2}\int_{-1}^1 dx\,
g(\mid\frac{{\bf r}}{2}+{\bf r}'\mid)g(\mid\frac{{\bf r}'}{2}+{\bf r}\mid),
\end{equation}
where $x$ denotes the cosine of the angle between ${\bf r}$ and ${\bf r}$.
We solved the radial eigenvalue equation (\ref{opera}) on a grid.

The lower bounds (\ref{lower}) for the three-boson energy 
$E_{0(3)}[\frac{3}{A}]$ can now be used, according to Eq.(\ref{patch}), 
to derive lower bounds ${\cal L}_0 \leq {\cal L}_1 \leq E_{0(A)}$
for the general $A$-boson system, 
\begin{eqnarray}
\label{eq:74}
{\cal L}_0 &=& \frac{A(A-1)}{6}g\left(\frac{2}{Amg}
\right)^{\frac{\beta}{\beta +2}}3\eta_{0(2)}=3C\eta_{0(2)}\nonumber\\
{\cal L}_1 &=& C\left(S_{0(2)}^{\mbox{\scriptsize max}}\eta_{0(2)}+
[3-S_{0(2)}^{\mbox{\scriptsize max}}]\eta_{1(2)}\right).
\end{eqnarray}
Note that, because of the scaling properties of power-law potentials, 
the $A$-dependence of ${\cal L}_0$ and ${\cal L}_1$ can be absorbed in the 
coefficient $C$ appearing in Eq.(\ref{eq:74}).

\begin{figure}
\centering{\
\psfig{file=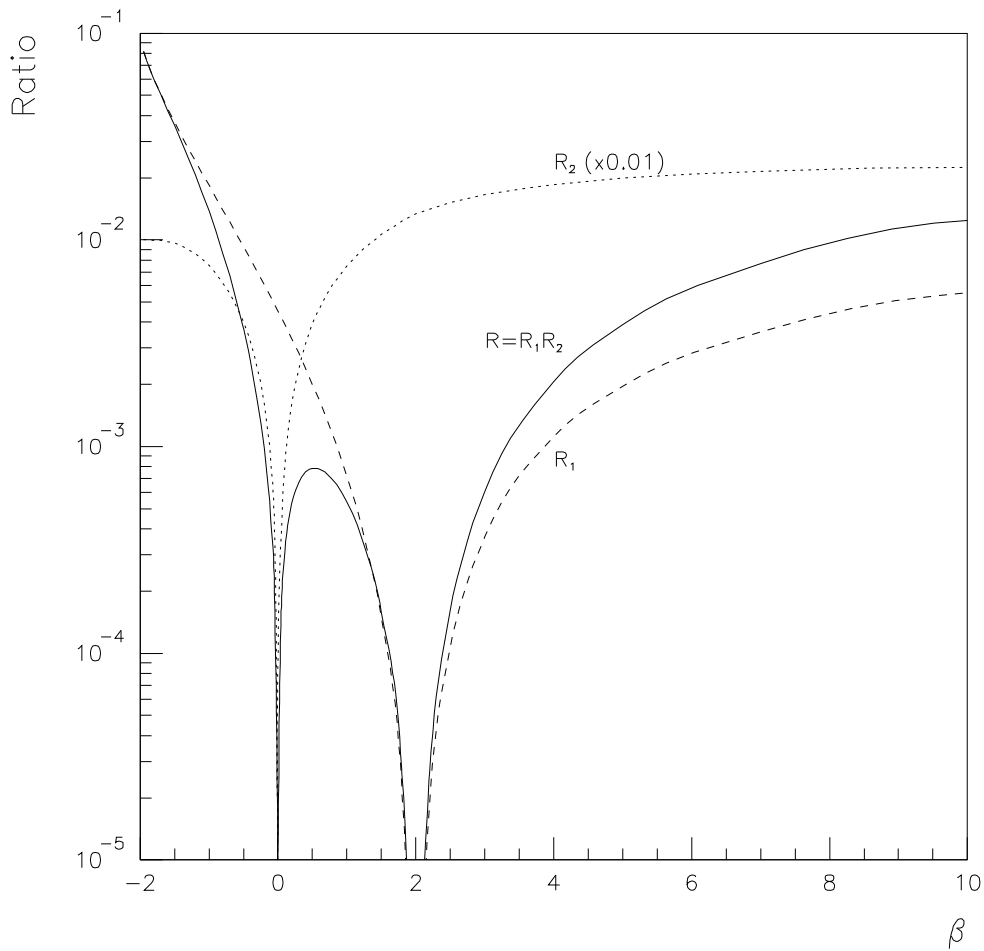,height=15cm}
\caption[fig5]{
Lower bounds for the energy of a system of $A$ bosons interacting with 
a power-law potential $\mbox{sgn}(\beta )r^{\beta}$. Solid line: 
relative improvement $R=({\cal L}_1 -{\cal L}_0 )/\mid{\cal L}_0 \mid$. 
The ratio $R=R_1 R_2$ is a product of two factors 
(see Eq.(\protect\ref{expl})), which are also plotted. Dashed line: $R_1 = 
1-S_{0(2)}^{\mbox{\scriptsize max}}/3$. Dotted line: 
$R_2 =(\eta_{1(2)}-\eta_{0(2)})/\mid\eta_{0(2)}\mid $.  
}
\label{fig5}
}
\end{figure}

In order to compare the new lower bound ${\cal L}_1$ with the 
conventional ${\cal L}_0$ we have plotted 
in Fig.~\ref{fig5} the relative improvement 
$R=({\cal L}_1 -{\cal L}_0 )/\mid{\cal L}_0 \mid $, for a range of 
powers $-2<\beta <10$. Being a ratio, $R$ is independent of $A$. It can 
be rewritten as a product
\begin{equation}
\label{expl}
R=\frac{{\cal L}_1 -{\cal L}_0}{\mid{\cal L}_0 \mid}=
\left(1-\frac{1}{3}S_{0(2)}^{\mbox{\scriptsize max}}\right)
\left(\frac{\eta_{1(2)}-\eta_{0(2)}}{\mid\eta_{0(2)}\mid}\right)=R_1 R_2,
\end{equation}
where the contributing factors $R_1$ and $R_2$ are related to the maximal 
occupation of the ground-state pair, and to the energy difference between 
ground and first excited state, respectively. These factors are also 
plotted in Fig.~\ref{fig5}. 

As can be seen from Fig.~\ref{fig5}, the relative improvement $R$ becomes 
zero for two values, $\beta=0$ and $\beta=2$. For $\beta=2$ it is $R_1$ that 
vanishes, since the operator in Eq.(\ref{opera}) has an eigenvalue equal to 
one if $g(r)\sim\exp{(-r^2)}$. This  reflects the fact that 
the conventional bound ${\cal L}_0$ becomes exact for harmonic oscillator 
systems\cite{Bas1-2}. For $\beta=0$ it is $R_2$ that vanishes, since 
the pair energy spectrum becomes degenerate 
as $\beta\rightarrow \pm 0$, that is, $\eta_{\nu (2)}\rightarrow \pm 1$ 
in this limit (see, e.g., \cite{hpnew}). 

Except for extreme values of $\beta$ (which means $\beta$ close to $-2$ or 
positive and large) the improvement of ${\cal L}_1$ over ${\cal L}_0$ seems 
modest, e.g.\ for the case of gravitating bosons ($\beta=-1$) we find 
$R\approx1.4\%$. However, for most power-law potentials the conventional 
bound ${\cal L}_0$ is already quite a good approximation to the exact energy 
of the three-body system, so any improvement cannot be large on this scale. 
In Table~1 we compare, for the three-body system with $m=g=1$, the exact 
ground-state energy (taken from \cite{Bas1-2}) with the lower bounds 
${\cal L}_0$ and ${\cal L}_1$, for a few values of $\beta$. It is seen that 
in the three-body system the improved bound removes a sizeable fraction 
(between $25\%$ and $75\%$) of the remaining discrepancy between the exact 
energy and the lower-bound of the Hall-Post type.  
 
It is interesting to note \cite{hpnew} that in the limit 
$\beta \rightarrow 0$ the power-law potential is related to the case of 
the logarithmic potential,  
$\ln{r} =\lim_{\beta\rightarrow 0}(r^{\beta}-1)/\beta $. As a consequence, 
the eigenvalues $\eta_{\nu}(\beta )$ of the hamiltonian 
$-{\bf \nabla}^2 +
\mbox{sgn}(\beta )r^{\beta}$ are, for small $\beta$, connected with the 
eigenvalues $\eta^L_{\nu}$ of the hamiltonian 
$-{\bf \nabla}^2 + \ln{r}$ via the 
relation 
\begin{equation}
\label{log} 
\eta_{\nu}(\beta )\rightarrow\mbox{sgn}(\beta )+\mid\beta\mid 
\left(\eta^L_{\nu} -\frac{1}{2}\ln{\mid\beta\mid}\right) , 
\end{equation}
which holds up to terms decreasing faster than linear in $\mid\beta\mid$. 
Eq.(\ref{log}) shows the origin of the degeneracy 
in the power-law eigenvalue spectrum for $\beta\rightarrow 0$. This 
degeneracy is absent if we 
consider directly the logarithmic potential, i.e.\ an $A$-boson system with 
hamiltonian 
\begin{equation}
\label{logham}
\hat{H}_A =-\frac{1}{2m}\sum_i {\bf \nabla}^2_i +g\,
\sum_{i_1 < i_2} \ln{\mid {\bf r}_i -{\bf r}_j \mid}.
\end{equation}
A straightforward analysis then leads to 
\begin{eqnarray}
{\cal L}_0 &=& \frac{A(A-1)}{6}g\left[3 \eta^L_{0(2)}+
\frac{1}{2}\ln{\left(\frac{2}{Amg}\right)}\right]\nonumber\\
{\cal L}_1 &=& \frac{A(A-1)}{6}g
\left[S_{0(2)}^{\mbox{\scriptsize max}}\eta^L_{0(2)}+
[3-S_{0(2)}^{\mbox{\scriptsize max}}]\eta^L_{1(2)}
+\frac{1}{2}\ln{\left(\frac{2}{Amg}\right)}\right], 
\end{eqnarray}
where $\eta^L_{0(2)}=1.0444332$, $\eta^L_{1(2)}=1.847442$, and 
$S_{0(2)}^{\mbox{\scriptsize max}} = 2.986419$.  
To compare this with the values in Fig.~\ref{fig5} one can consider, e.g.,   
$mAg/2\approx 1$. The relative improvement then becomes 
$R=\frac{{\cal L}_1 -{\cal L}_0}{\mid{\cal L}_0 \mid}\approx 0.35\%$.  
 
\subsection{Electrons confined in a harmonic oscillator 
well.}\label{section6.3}
As an example of a fermion problem we improve the bounds derived recently 
by Juillet et al.\ for a quantum dot system of electrons confined in a 
harmonic oscillator well\cite{Jui}. The hamiltonian in atomic units ($m=e=1$) reads
\begin{equation} 
\hat{H}=\sum_{i=1} \left( -\frac{{\bf \nabla}_i^2}{2}
+\frac{\omega^2{\bf r}_i^2}{2}\right)+\sum_{i<j}
\frac{1}{\mid{\bf r}_i -{\bf r}_j \mid^2} .
\end{equation}
The harmonic center-of-mass motion can be split off and treated exactly, and 
we concentrate on the relative motion. 

We consider a system of three electrons with total spin $S=\frac{3}{2}$, 
e.g.\ three spin-up electrons. The spatial wave function is antisymmetric, 
so the bound derived in Section~\ref{section5} can be applied for the energy 
of the ground state, which has $L^{\pi}=1^+$. 

According to Eq.(\ref{eq:7}), with $N=2$ and $A=3$, we must first construct 
the ground state of the relative two-body hamiltonian,
\begin{equation}
\label{eq:71}
{\hat{H}'}_2=\frac{2}{3}\left(-{\bf \nabla}^2 +\frac{\omega^2}{4} 
{\bf r}^2 \right) +\frac{1}{r} .
\end{equation}
In the previous examples the pair ground state was nondegenerate.  
In the present case the lowest antisymmetric eigenstate $g({\bf r})$ forms 
a $L^{\pi}=1^-$ triplet,
\begin{equation}
g_{\mu}({\bf r}) =g(r)\mbox{Y}_{1\mu}(\Omega).
\end{equation}
Since the pair ground state is now degenerate, we must 
generalize Eq.(\ref{eq:59}), and maximize the {\em joint} occupancy   
$\lambda =\sum_{\mu}S_{(2)\mu}$ of the members of the triplet. 
This leads in a straightforward fashion to an eigenvalue equation,
\begin{equation}
\label{eq:73}
(\lambda-1)G_{\mu\nu}({\bf r}) =-2\sum_{\mu'}\int d{\bf r}' 
\left[g^*_{\mu}(\frac{{\bf r}}{2}+{\bf r}')
g_{\mu'}(\frac{{\bf r}'}{2}+{\bf r})\right] G_{\mu'\nu}({\bf r}'),
\end{equation}
which replaces Eq.(\ref{neweq}). The overlap function $G_{\mu\nu}({\bf r})$ 
between the $1^-$ pair state $g_{\mu}$ and one of the members 
$\Psi_{(3)\nu}$ of the $A=3$ ground-state $1^+$ triplet, has the following 
tensor structure,
\begin{equation}
G_{\mu\nu}({\bf r}) =G(r)
\langle\, 1\, \mu\, 1\, \nu-\mu\mid\,1\, \nu\rangle 
\mbox{Y}_{1\nu-\mu}(\Omega).
\end{equation}
Substitution into Eq.(\ref{eq:73}) leads, after some angular momentum 
algebra, to a  radial eigenvalue equation 
\begin{equation}
\label{eq:75}
(\lambda -1)G(r)=2\int dr' {r'}^2 W(r,r')G(r').
\end{equation}  
The operator $W$ reads  
\begin{equation}
W(r,r')=\frac{1}{2}rr'\int_{-1}^1 dx\,
\left(\mbox{P}_0 (x)-\mbox{P}_2 (x)\right)
\frac
{g(\mid\frac{{\bf r}}{2}+{\bf r}'\mid)g(\mid\frac{{\bf r}'}{2}+{\bf r}\mid)}
{\mid\frac{{\bf r}}{2}+{\bf r}'\mid\mid\frac{{\bf r}'}{2}+{\bf r}\mid},
\end{equation}
where $\mbox{P}_l$ are the Legendre polynomials. 

We checked that for a HO $p$ wave function, $g(r)\sim r\exp{(-r^2)}$, the 
maximal 
eigenvalue of Eq.(\ref{eq:75}) yields $\lambda=3$, which equals the number of 
pairs in 
the $A=3$ system. This means that the Hall-Post bound ${\cal L}_0$ already 
coincides with the exact result for HO systems, as could be expected from the 
discussion in \cite{Jui}.

In the presence of the electron-electron repulsion the pair wave function 
is distorted from the HO shape, and we find a maximal eigenvalue 
$\lambda =2.901644$ for $\omega=0.01$. The lowest energies of the 
hamiltonian (\ref{eq:71}) are ${\epsilon}_0 =0.05560837$ for the 
($p$-wave) ground state and   
${\epsilon}_1 =0.06171271$ for the ($f$-wave) first excited antisymmetric 
state. For the Hall-Post bound we thus find 
${\cal L}_0 =3{\epsilon}_0 = 0.1668$, in agreement 
with \cite{Jui}. This is already quite a good bound compared 
with the exact three-body ground-state energy $E_{(3)}= 0.1680$, as quoted 
in \cite{Jui}. The new bound improves this to ${\cal L}_1 =\lambda\epsilon_0 +
(3-\lambda )\epsilon_1 =0.1674$\,. 

For $\omega=10$, which is closer to a pure harmonic oscillator system, we 
find $\lambda=2.999780$, $\epsilon_0 =18.32273$, and $\epsilon_1 =31.14720$, 
yielding bounds ${\cal L}_0 =54.968$ and  ${\cal L}_1 =54.971$, to be 
compared with the exact result $E_{(3)}=54.973$, quoted in \cite{Jui}.

In conclusion, by taking the structure of the pair wave function into account 
we are able to halve the remaining deviation between the Hall-Post 
type lower bound and the exact result. 

\section{Summary}\label{section7}

Motivated by the renewed interest in lower 
bounds for the ground-state energy of many-body systems, we have developed 
a method to improve the existing lower bounds of the Hall-Post type. 
The method is based on an exact sumrule for the energy in terms of 
two-body occupation numbers (or, equivalently, spectroscopic factors related 
to $A-2$-particle removal) in the $A$-particle ground state. The pair 
occupation numbers that enter the sumrule refer to the two-body eigenstates 
of the two-body cluster hamiltonian in the conventional 
Hall-Post decomposition of the many-body hamiltonian. We find that it is 
possible to derive upper bounds for these pair occupation numbers, without 
detailed knowledge of the structure of the $A$-particle wave function. 
These upper bounds, or maximal pair occupancies, do depend on the structure 
of the pair state, and can be used to obtain strict lower bounds to the 
$A$-particle energy which are better than the conventional one. 

We have studied both the bosonic and fermionic sector, and developed a 
framework for both fixed-center systems and self-bound systems, where the 
wave functions are translationally invariant. We have applied the formal 
results to various numerical examples, and demonstrated that significant 
improvements are obtained over the conventional lower bound. 

Several problems are still remaining. In the case of self-bound systems 
a method to evaluate the maximal pair occupation in $A$-particle space 
is not available for $A\geq 4$. Also states of mixed spatial symmetry 
are not yet treated. If the two-body energy spectrum contains a continuum 
part, the associated pair strength does not contribute to the 
lower bound in the present work; a better treatment of the continuum part 
would be very 
interesting, as it would lead to the derivation of improved bounds for the 
critical coupling strength \cite{Ric,Goy,Mos} needed to achieve binding in 
many-body systems. 

In general, further improvements could be made by more 
refined approximations to the distribution $S_{\nu (2)}$ of the pair 
occupations over the various pair states. In the present work this 
occupation is maximized for each pair state separately; in reality, of 
course, they are interrelated, as they reflect occupations within the same 
$A$-particle state. Such a refinement is in particular needed for the 
fermion case, since the noninteracting limit is at present not reproduced. 

The present work can also be rephrased in terms of abstract many-body 
theory. The $A$-particle ground-state energy in 
Eqs.(\ref{eq:18},\ref{eqself}) is expressed as 
\begin{equation}
\label{final}
E_{0(A)}=\mbox{Trace\,}\{\hat{H}_2 [\alpha]\rho^{(2)}\},
\end{equation}
where $\alpha =1/(A-1)$ or $\alpha =2/A$ for fixed-center and self-bound 
systems, respectively. Minimizing the r.h.s\ of Eq.(\ref{final}) over all 
$A$-representable two-body densities would yield the exact $A$-particle 
energy. The full set of exact conditions for $A$-representability are of 
course unknown. Minimizing the r.h.s.\ of Eq.(\ref{final}) over all 
two-body densities which comply with a {\em limited} set of 
$A$-representability conditions will then yield a lower bound for the 
$A$-particle energy. The conventional bound ${\cal L}_0$ can be seen 
as the lowest-order approximation in this scheme, since only the 
normalization condition $\mbox{Trace\,}\{\rho^{(2)}\}=A(A-1)/2$ is required 
for the two-body density matrix in Eq.(\ref{final}). The improved bounds 
in this work can be viewed as imposing  additional conditions on the natural 
pair occupation numbers of the two-body density matrix in this scheme.   

This work was supported by the Fund for Scientific Research-Flanders 
(FWO-Vlaanderen) and the Research Council of Ghent University. 

\begin{appendix}
\section{Natural orbital representation for pair states}\label{app1}
\subsection{Two-boson states}\label{app1.1}
In a general SP basis, a two-boson state can be expanded as 
\begin{equation}
\mid\Psi_{(2)}\rangle =\frac{1}{\sqrt{2}}\sum_{\alpha\beta}
C_{\alpha\beta}c_{\alpha}^{\dagger}c_{\beta}^{\dagger}, 
\end{equation}
where $C_{\alpha\beta}=C_{\beta\alpha}$ is a (complex) symmetric matrix, and 
$\mbox{Trace}\{CC^{\dagger}\}=\sum_{\alpha\beta}\mid C_{\alpha\beta}\mid^2 =1$ 
fixes the normalization.

The hermitian one-body density matrix reads
\begin{equation}
\rho_{\alpha\beta}=\langle\Psi_{(2)}\mid c_{\beta}^{\dagger}
c_{\alpha}\mid\Psi_{(2)}\rangle=
2\sum_{\lambda}C_{\alpha\lambda}C^*_{\beta\lambda}, 
\end{equation} 
or, in matrix notation, $\rho=2CC^{\dagger}$.

Under a unitary transformation 
$c'^{\dagger}_{\alpha'} =\sum_{\alpha}U_{\alpha\alpha'}
c_{\alpha}^{\dagger}$ to a new SP basis, 
the matrices $\rho$ and $C$ transform as 
\begin{equation}
\label{a4}
\rho' =U^{\dagger}\rho U,\,\,\,\,\,\, C'=U^{\dagger} C U^* . 
\end{equation}

We can always make a unitary transformation to the natural SP basis that 
diagonalizes the one-body density matrix $\rho$. In this basis $\rho$ is 
real, and as a 
consequence the commutators $[C,C^{\dagger}]=[\rho ,C] =0$ vanish. 
It follows that in the natural basis $C$ is block diagonal, each $n\times n$ 
block $C^{(\alpha)}$ corresponding to a $n$-fold degenerate 
$\rho_{\alpha\alpha}$. 

For such a block the matrix 
$D^{(\alpha)}=C^{(\alpha)}\sqrt{2/\rho_{\alpha\alpha}}$ is a unitary 
and symmetric matrix, which can be diagonalized by a real orthogonal 
transformation. Indeed, if $X$ is an eigenvector of $D^{(\alpha)}$ with 
eigenvalue $\lambda$, then $D^{(\alpha)} X =\lambda X$ implies 
$D^{(\alpha)} X^* =\lambda X^*$, because $D^{(\alpha)^*} 
=D^{(\alpha)^{-1}}$ and 
$\lambda^* =\lambda^{-1}$. So either $X=X^*$ is real, or $\lambda$ is 
degenerate with eigenvectors $X,X^*$, which can be replaced by an orthogonal 
pair from the real linear combinations $X_+ =(X+X^* ), X_- =(X-X^* )/i$. 

Since the transformation to the basis of real eigenvectors of $C$ is 
real orthogonal, it also corresponds, according 
to Eq.(\ref{a4}), to an allowed unitary transformation on the SP basis. 
In this basis, 
$C^{(\alpha)}_{mn}=\delta_{mn}
\sqrt{\rho_{\alpha\alpha}/2}\exp(i\theta_n )$. The phase can be absorbed 
in the SP states. As a result, the desired canonical form of a two-boson 
state reads,
\begin{equation}
\mid \Psi_{(2)}\rangle =\frac{1}{\sqrt{2}}
\sum_{\alpha}x_{\alpha}(c^{\dagger}_{\alpha})^2
\mid \rangle ,
\end{equation}
where $x_{\alpha}=\sqrt{\rho_{\alpha\alpha}/2}$ is real and positive.

\subsection{Two-fermion states}\label{app1.2}

In a general SP basis, a two-fermion state can be expanded as 
\begin{equation}
\mid\Psi_{(2)}\rangle =\frac{1}{\sqrt{2}}\sum_{\alpha\beta}
C_{\alpha\beta}
c_{\alpha}^{\dagger}c_{\beta}^{\dagger}\mid\rangle, 
\end{equation}
where $C_{\alpha\beta}=C_{\beta\alpha}$ is a (complex) antisymmetric matrix, 
and $\mbox{Trace}\{CC^{\dagger}\}=1$.
 
An identical analysis as in the bosonic case leads in the natural basis 
to matrices $D^{(\alpha)}$ which are now unitary and antisymmetric, and can 
be brought to a canonical form by a real orthogonal transformation.
If $D^{(\alpha)}X =\lambda X$, then $D^{(\alpha)} X^* =-\lambda X^*$ 
(because $D^{(\alpha)^*} =-D^{(\alpha)^{-1}}$). So the 
eigenvalues/vectors come in pairs 
$(\lambda ,X),(-\lambda ,X^*)$, where $X^{\dagger}X^* =0$. For such a pair 
we may transform from the eigenvector basis $X,X^*$ to the real basis 
$X_+ =(X+X^* )/\sqrt{2},X_- =(X-X^* )/i\sqrt{2}$. The  
$2\times 2$ diagonal block in the $X,X^*$ representation is 
transformed as
\begin{equation}
\left(
\begin{array}{cc}\lambda & 0 \\ 0 & -\lambda \end{array}\right)
\rightarrow
\left(
\begin{array}{cc}0 & -i\lambda \\ i\lambda & 0 \end{array}\right) .
\end{equation}

The transformation to the real basisvectors $X_+ ,X_- $ is again 
real orthogonal, and corresponds to 
an allowed unitary transformation on the SP basis. In this basis, 
$C^{(\alpha)}_{mn}=\delta_{m\bar{n}}
\sqrt{\rho_{\alpha\alpha}/2}\exp(i \theta_m)$, where $n,\bar{n}$ are  
associated pair states and $\theta_{\bar{n}} =\theta_n +\pi$. 
Apart from an overall sign, the phase can be absorbed 
in the SP states. As a result, the desired canonical form of a two-fermion 
state reads,
\begin{equation}
\mid \Psi_{(2)}\rangle =\frac{1}{\sqrt{2}}
\sum_{\alpha > 0}x_{\alpha}(
c^{\dagger}_{\alpha}c^{\dagger}_{\bar{\alpha}}-
c^{\dagger}_{\bar{\alpha}}c^{\dagger}_{\alpha})
\mid \rangle =
\sqrt{2}\sum_{\alpha > 0}x_{\alpha}
c^{\dagger}_{\alpha}c^{\dagger}_{\bar{\alpha}}
\mid \rangle ,
\end{equation}
where $x_{\alpha}=\sqrt{\rho_{\alpha\alpha}/2}$ is real and positive, and 
the summation $\alpha >0$ is made over distinct pairs.

\section{Maximal pair occupancy for a general number of bosons}\label{app2}

We follow here closely the reasoning by Pan et al.\ \cite{Pan} for the fermion 
pairing problem.

A two-boson state $\Psi_{(2)}$ is expressed in its natural basis as 
\begin{equation}
\mid \Psi_{(2)}\rangle =\frac{1}{\sqrt{2}}\sum_{\alpha}
x_{\alpha}(c^{\dagger}_{\alpha})^2\mid\rangle =\varphi_{(2)}^{\dagger}
\mid\rangle ,
\end{equation} 
where the $x_{\alpha}$ are real and positive, and 
$\sum_{\alpha}x_{\alpha}^2 =1$. The construction of the eigenvalues 
of the $\varphi^{\dagger}_{(2)}\varphi^{\mbox{}}_{(2)}$ operator then  
proceeds as follows.  

The uncorrelated $A$-boson states in the natural SP basis read 
\begin{equation}
\mid\Psi_{(A)}\rangle = \prod_{i=1}^{+\infty}(c^{\dagger}_i )^{p_i}
\mid\rangle , 
\end{equation}
where $\sum_i p_i =A$, and 
can be classified according to the broken pairs they contain, i.e.\ with 
$p_i =2m_i +r_i$ and $r_i =0$ or $r_i =1$, we can construct the corresponding 
vacuum $\mid 0\rangle$ of the $\varphi_{(2)}$ operator, 
\begin{equation}
\mid 0\rangle = \prod_{i=1}^{+\infty}(c^{\dagger}_i )^{r_i}\mid\rangle .
\end{equation}
Obviously, the $\varphi^{\dagger}_{(2)}\varphi^{\mbox{}}_{(2)}$ operator 
does not connect $A$-boson states belonging to a different vacuum 
$\mid 0\rangle$. For our purpose we may take $\mid 0\rangle =\mid\rangle$ 
(the zero-boson state) if $A=2a$ is even, and  $\mid 0\rangle =
c^{\dagger}_{\beta}\mid\rangle =\mid\beta\rangle$ if $A=2a+1$ is odd, where 
$\beta$ is one of the single-boson states.    

Next we introduce generalized operators 
\begin{equation}
\varphi^{\dagger}(y)=\frac{1}{\sqrt{2}}
\sum_{\alpha}\frac{x_{\alpha}}{1-yx_{\alpha}^2}
(c_{\alpha}^{\dagger})^2;\;\;\;
N(y)=\sum_{\alpha}\frac{x_{\alpha}^2}{1-yx_{\alpha}^2}
\left(c_{\alpha}^{\dagger}c_{\alpha}^{\mbox{}}+\frac{1}{2}\right) ,
\end{equation}
which obey the commutator algebra
\begin{equation}
\label{alg}
\left[ \varphi^{\mbox{}}(y_1 ),\varphi^{\dagger}(y_2 )\right] 
=2\frac{y_1 N(y_1 )-y_2 N(y_2 )}{y_1 -y_2 };\;\;\;
\left[ N(y_1 ), \varphi^{\dagger}(y_2 )\right] =
2\frac{\varphi^{\dagger}(y_1 )-\varphi^{\dagger}(y_2 )}{y_1 -y_2} .
\end{equation}

One can now show that for a suitable choice of variables 
$y_1 ,\dots ,y_{a-1}$, the vector 
\begin{equation}
\mid\Psi_{(A)}\rangle =\varphi^{\dagger}(0)
\varphi^{\dagger}(y_1 )\dots\varphi^{\dagger}(y_{a-1})\mid 0\rangle ,
\end{equation}
is an eigenvector of the $\varphi^{\dagger}(0)\varphi (0)=
\varphi^{\dagger}_{(2)}\varphi^{\mbox{}}_{(2)}$ operator.

Using the commutation relations (\ref{alg}), one finds 
\begin{eqnarray}
\label{long}
\varphi^{\dagger}(0)\varphi (0)\mid\Psi_{(A)}\rangle &=& 
\left(2\Lambda (0) +4\sum_{k=1}^{a-1}\frac{1}{y_k}\right)
\mid\Psi_{(A)}\rangle \\
&&
+2\sum_{k=1}^{a-1}\left(\Lambda (y_k )-\frac{2}{y_k}
-\sum_{\tiny \begin{array}{cc}k'=1\\ 
(k'\neq k)\end{array} }^{a-1}
\frac{2}{y_k -y_{k'}}\right)\varphi^{\dagger}(0)\varphi^{\dagger}(0)
\prod_{\tiny \begin{array}{cc}i=1\\(i\neq k)\end{array} }^{a-1}
\varphi^{\dagger}(y_i )\mid 0\rangle , \nonumber 
\end{eqnarray}
where $\Lambda (y)$ is the eigenvalue for $N (y)$ acting on the vacuum 
$\mid 0\rangle$,   
\begin{equation}
\Lambda (y) =\sum_{\alpha}\frac{x_{\alpha}^2}{1-yx_{\alpha}^2}
\left(\frac{1}{2}+s\delta_{\alpha\beta}\right) .
\end{equation}
For even $A$, $\mid 0\rangle =\mid\rangle$ and $s=0$; for odd $A$, 
$\mid 0\rangle =\mid\beta\rangle$ and $s=1$. 

The second term in Eq.(\ref{long}) vanishes if the variables 
$(y_1 , \dots ,y_{a-1})$ are solutions of the set (\ref{difficult}) 
of nonlinear equations, whereas the coefficient in front of the first term 
yields the corresponding eigenvalue (\ref{eigenvalue}) of the 
$\varphi^{\dagger}(0)\varphi (0)$ operator.
\end{appendix}

\begin{table}[table1]
\caption{Energies and bounds in a three-body system 
interacting with power-law potentials $\mbox{sgn}(\beta )r^{\beta}$. 
The hamiltonian 
is given by Eq.(\protect\ref{powlawham}) with $A=3$ and $m=g=1$. We compare 
the lower bounds ${\cal L}_0$ and ${\cal L}_1$ with the exact energy 
$E_{0(3)}$, taken from \protect\cite{Bas1-2} and properly rescaled. The last 
column contains the ratio $({\cal L}_1-{\cal L}_0 )/(E_{0(3)}-{\cal L}_0 )$.}
\begin{center}
\begin{tabular}{|r|r|r|r|r|}
 $\beta$ & $E_{0(3)}$ & ${\cal L}_0$ &${\cal L}_1$& ratio\\
\hline
-1.0 & -1.0670 & -1.1250 & -1.1095 & 27 \\
-0.5 & -1.4911 & -1.5043 & -1.4987 & 42 \\
0.1 & 3.6383 & 3.6363 & 3.6374 & 57 \\
0.5 & 5.0780 & 5.0718 & 5.0757 & 64 \\
1.0 & 6.1323 & 6.1276 & 6.1309 & 71 \\
2.0 & 7.3485 & 7.3485 & 7.3485 & - \\
3.0 & 8.1228 & 8.1163 & 8.1212 & 75 \\
\end{tabular}
\end{center}
\end{table}

\end{document}